\documentclass{jfm}
\usepackage{graphicx}
\usepackage{epstopdf, epsfig}
\usepackage[colorlinks,citecolor = blue, linkcolor=blue,hyperindex,CJKbookmarks]{hyperref}
\usepackage{siunitx}
\usepackage{amsmath}

\newcommand{\reviewerA}[1]{\textcolor{black}{#1}}
\newcommand{\reviewerB}[1]{\textcolor{black}{#1}}
\newcommand{\reviewerC}[1]{\textcolor{black}{#1}}

\newcommand{\software}[1]{\textsc{#1}}
\newcommand{\CA}{$\theta$}
\newcommand{\MAT}{\operatorname{Ma}_T}
\newcommand{\MAG}{\operatorname{Ma}_\Gamma}
\newcommand{\RAT}{\operatorname{Ra}_T}

\shorttitle{Solutal Marangoni effects overwhelm thermal Marangoni flow}
\shortauthor{D. Rocha, P. L. Lederer, P. J. Dekker, A. Marin, D. Lohse and C. Diddens}

\title{Evaporating sessile droplets: solutal Marangoni effects overwhelm thermal Marangoni flow}

\author{
Duarte Rocha\aff{1} \corresp{\email{d.rocha@utwente.nl}},
Philip L. Lederer\aff{2},
Pim J. Dekker \aff{1},
Alvaro Marin\aff{1},
Detlef Lohse\aff{1,3} \corresp{\email{d.lohse@utwente.nl}},
\and Christian Diddens\aff{1} \corresp{\email{c.diddens@utwente.nl}}
}

\affiliation{\aff{1}Physics of Fluids Department, Max-Planck Center Twente for Complex Fluid Dynamics and J. M. Burgers Centre for Fluid Dynamics, University of Twente, 7500AE Enschede, The Netherlands,
\aff{2}Department of Mathematics, Universit\"at Hamburg,
Bundesstr. 55, 20146, Hamburg, Germany
\aff{3}Max-Planck Institute for Dynamics and Self-Organization, Am Faßberg 17, 37077 Göttingen, Germany
}

\begin{document}

\maketitle

\begin{abstract}
When an evaporating water droplet is deposited on a thermally conductive substrate, the minimum temperature will be at the apex due to evaporative cooling.
Consequently, density and surface tension gradients emerge within the droplet and at the droplet-gas interface, giving rise to competing flows from, respectively, the apex towards the contact line (thermal-buoyancy-driven flow) and the other way around {(thermal Marangoni flow)}. 
In small droplets with a diameter below the capillary length, the thermal Marangoni effects are expected to dominate over thermal buoyancy (``thermal Rayleigh'') effects. 
However, contrary to these theoretical predictions, our experiments mostly show a dominant circulation from the apex towards the contact line, indicating a prevailing of thermal Rayleigh convection. 
Furthermore, our experiments often show an unexpected asymmetric flow that persisted for several minutes. 
We hypothesise that a tiny amount of contaminants, commonly encountered in experiments with water/air interfaces, act as surfactants and counteract the thermal surface tension gradients at the interface and thereby promote the dominance of Rayleigh convection. 
Our finite element numerical simulations demonstrate that, under our specified experimental conditions, a mere 0.5\% reduction in the static surface tension caused by surfactants leads to a reversal in the flow direction, compared to the theoretical prediction without contaminants. 
Additionally, we investigate the linear stability of the axisymmetric solutions, revealing that the presence of surfactants also affects the axial symmetry of the flow.
\end{abstract}


\section{Introduction}\label{sec:introduction}
Evaporation of droplets is a phenomenon present in many industrial applications, such as inkjet printing \citep{lohse2021fundamental,mishra2010high, herman2018drop, perelaer2009droplet}, spray drying \citep{perdana2011single,vehring2007particle,eslamian2009modeling} or pesticide administration \citep{yu2009evaporation, yu2009evaporation2,luo1994characteristics}.
Understanding the characteristics of the flow inside evaporating droplets is essential to predict the final deposition patterns, which is of great importance in the aforementioned applications. 
Industrial processes often involve fluids with complex rheological properties \citep{deGans2004inkjet} or with multiple components \citep{kim2016controlled, tan2016evaporation, diddens2017modeling, edwards2018density, li2019gravitational, lohse2020physicochemical, wang2022wetting}, but the study of the evaporation of pure water droplets is an obvious first step towards the comprehension of more intricate systems.

Water droplets are ubiquitous in nature and their evaporation is seen daily on e.g.\ raindrops on a leaf, drying dishes on a rack, or dew on a spider web. 
In any of the latter examples, the droplet generally evaporates at ambient conditions. 
As the early investigations of \citet{deegan1997capillary,deegan2000contact,deegan2000pattern} and \citet{popov2005evaporative} showed, the non-uniform evaporation rate at the liquid-gas interface can be explained as a consequence of the diffusion-limited transport of water vapour away from the interface into the far-field atmosphere.
When the droplet is sitting on a hydrophilic substrate, i.e. with a contact angle \CA\ lower than $90^{\circ}$, the evaporation rate is higher close to the contact line, while for a hydrophobic substrate (\CA\ $>90^{\circ})$, the evaporation rate is maximum at the apex \citep{WilsonANRFM2023}.
In combination with the contact line dynamics, diffusion-limited evaporation leads to intriguing phenomena like the ``coffee-stain effect'' \citep{deegan1997capillary}, where solutes are transported towards the pinned contact line through a capillary-driven flow, creating distinct ring-shaped deposition patterns.
Evaporation-induced cooling at the droplet-gas interface, owing to latent heat release \citep{schreiber1981conductive,ward2001interfacial,hu2002evaporation,WilsonANRFM2023,gelderblom2022evaporation}, also impacts the evaporation dynamics by establishing temperature gradients along the interface. 
A locally higher evaporation rate at the contact line, if \CA\ $< 90^{\circ}$, or at the apex, if \CA\ $> 90^{\circ}$, induces lower temperature in these regions.
However, if the thermal conductivity of the substrate is high enough, the substrate will remain close to the ambient temperature $T_0$, and the distance to the substrate becomes the most relevant factor influencing the temperature distribution within the droplet \citep{ristenpart2007, dunn2009strong, girard2006evaporation,shahidzadeh2006evaporating}. 
The temperature will then be minimal at the apex.
Temperature differences along the interface of the droplet induce a surface tension gradient, which in turn drives a significant thermal Marangoni flow \citep{sultan2004evaporation}, typically from the contact line towards the apex \citep{ristenpart2007,dunn2009strong}.
Additionally, a buoyant (``thermal Rayleigh'') flow from the apex to the contact line might also appear \citep{savino2002buoyancy}, driven by the density gradient in the bulk of the droplet \citep{diddens_li_lohse_2021,gelderblom2022evaporation}.

Even though theoretical predictions \citep{hu2005analysis,saenz2015evaporation,diddens2017evaporating} suggest that there should be a strong thermal Marangoni flow in evaporating pure water droplets sitting on highly thermally conductive substrates, experiments \citep{buffone2004investigation, marin2016surfactant, diddens2017evaporating,rossi2019interfacial} have often shown that there is only a weak (or no) presence of such effect.
Discrepancies between experimental results and theoretical predictions in this context are commonly ascribed to the presence of contaminants \citep{savino2002buoyancy, hu2005analysis,hu2006marangoni, girard2008effect, xu2007marangoni, van2022competition} that act as surfactants. 
While the specific nature of these impurities remains unknown \citep{gelderblom2022evaporation}, their existence at the interface between water and air has been identified in previous observations \citep{molaei2021interfacial,rey2018dirty} and surface tension at these interfaces has been observed to considerably decrease over time \citep{montanero2016ponce}, indicating surface aging due to an adsorption process. 

\citet{rossi2019interfacial} showed that, even though strong ionic concentration gradients in mineral-laden water droplets can lead to a reduction of thermal Marangoni flow, the velocity magnitude inside evaporating ultrapure (deionised) and common spring (mineral) water droplets is of the same order, suggesting that minerals are not the only source of a generally present contaminant in water droplets.
Exceptionally, strong thermal Marangoni flows have been experimentally observed by \citet{kazemi2021marangoni} in evaporating high contact angle droplets of ultrapure, degassed water on copper rods.
These authors registered velocity magnitudes up to \SI{10}{\milli\meter\per\second}, which is of the same order of magnitude as their numerical predictions.
However, the authors emphasise that after a short time period, the observed thermal Marangoni vortices become weaker, ultimately disappearing, therefore corroborating the hypothesis of contaminants increasingly acting on the surface. 
\citet{van2022competition} numerically co-validated the results from the lubrication theory and from quasi-stationary Stokes flow models to show the influence of both insoluble and soluble surfactants in strongly reducing the thermal Marangoni flow in evaporating droplets at ambient conditions. 
Thermal Rayleigh forces were disregarded due to their weak contribution to the flow, which is typically valid when considering small contact angles and non-heated substrates \citep{gelderblom2022evaporation}. 

Buoyant forces can, however, be relevant for the flow features in droplets with large contact angle.
These can lead to intriguing effects such as breaking of the axial symmetry the flow, beyond some critical Rayleigh number.
Experimentally, such symmetry breaking has been observed in e.g.\ small spherical Leidenfrost droplets \citep{bouillant2018leidenfrost,yim2022leidenfrost} or in evaporating droplets on heated hydrophobic \citep{josyula2021insights} and superhydrophobic \citep{tam2009marangoni,dash2014buoyancy,peng2020non} substrates.
\citet{dash2014buoyancy} suggests that the breaking of the axial symmetry into a single vortex within large contact angle droplets is correlated with the dominance of buoyancy-induced flow, analogous to the non-axial-symmetric Rayleigh-Bénard convection in cylindrical containers of aspect ratio 1 \citep[see][]{shishkina2021rayleigh}.

In this work, we present new experimental observations on the evaporation of high contact angle water droplets. 
Our findings provide further evidence of the suppression of thermal Marangoni flow, with thermal Rayleigh flow becoming the dominant mechanism. 
Additionally, in some experiments, we observed the breaking of the axial symmetry, leading to the formation of a single vortex early in the evaporation process, ultimately transitioning to an axisymmetric and buoyancy-dominated flow.
Throughout this paper, our main focus is to explain through numerical simulations the behaviours reported in our experiments.
For that purpose, we use a sophisticated hybrid discontinuous Galerkin (HDG) and finite element method (FEM) to investigate the influence of surfactants on the flow direction and the azimuthal stability of the flow inside an evaporating water droplet and compare with the experimental results. 

This paper is structured as it follows: In \S\ref{sec:experiments}, we report our new experimental observations. 
In \S\ref{sec:governing_equations}, the governing equations of the transient and quasi-stationary numerical models are presented, as well as the numerical procedure.
In \S\ref{sec:pure_water_droplet}, numerical predictions for flow direction in a pure water droplet are compared to the experimental results.
A comprehensive analysis in $\MAT-\RAT$ phase diagrams of the flow direction (assuming axisymmetry) and of the azimuthal stability of the stationary solutions is then presented for pure water droplets.
Contaminants are considered in \S\ref{sec:contaminants}, where we first introduce the additional equations and parameters for the insoluble surfactants.
We then follow up by showing the influence of surfactants on the flow direction at the experimental conditions.
Afterwards, we take different $\MAG$ values to perform the same phase space analysis as for pure water droplets.
Finally, in \S\ref{sec:conclusions} we end the paper with a summary, main conclusions, and an outlook.

\section{Experiments}\label{sec:experiments}

\subsection{Experimental setup}\label{sec:experimental_setup}


Experiments were performed on sessile droplets deposited on a superhydrophobic substrate, consisting of a micro-structured glass slide coated with a fluorinated monolayer to increase its hydrophobicity. This yielded static macroscopic contact angles larger than 150$^\circ$ and very low roll-off angles (low hysteresis). 
The flow inside the droplets was visualised by fluorescently labelled polystyrene particles (provided by \textit{microParticles GmbH}) with a diameter of approximately \SI{1}{\micro\meter}, which makes them suitable as flow tracers. Such colloidal particles are stabilised by sulfate groups at their surface, which conveys stability to the suspension with no need of surfactants, which would interfere with our measurements.
The flow was visualised and quantified by particle image velocimetry (PIV), for which we used a very low particle concentration ($5\cdot10^{-5} \ w/v\%$). Illumination was performed by a thin laser sheet, aimed at the central cross-section of the droplet. Imaging was performed with Ximea CCD camera, mounted on a long-distance microscope, at a frame rate of 0.75 frames per second, which was enough to capture the slow motion of the particles. PIV images were processed using our own algorithm. We used a multi-pass process with decreasing window size, from 128$\times$128 pixels down to 32$\times$32 pixels, with a 50\% window overlap in each pass. 

Due to the spherical-cap shape of the droplet, the light coming from the particles into the sensor experience refraction at the liquid-air interface and therefore the particles appear at different positions in the image than they are in the drop. After the images had been processed with the PIV algorithm, we applied an optical correction using ray-tracing to the centres of the interrogation windows and the measured displacements. Velocities that were measured close to the edges of the drop ($r/R>0.9$) were omitted, since there the optical distortion becomes very large, leading to unreliable results.

The droplet was carefully deposited over the superhydrophobic substrate inside a closed (not pressurised) chamber. \reviewerA{The substrate was previously cleaned by rinsing it once in water and carefully drying it with a nitrogen gas flow. To reduce the impact of contamination from the air in the room, the chamber was cleaned with a wet cloth and it remained then closed until the experiment was performed.} The temperature inside the chamber was not controlled, but remained constant at 21 $^\circ$C throughout the whole experiment. The relative humidity was passively controlled using reservoirs of pure water or salt-saturated water solutions in the chamber, which allowed us to vary the relative humidity from $35\%$ to $90\%$. The relative humidity and temperature were measured inside the chamber using a sensor (HGC 30 from DataPhysics) near the droplet.

\subsection{Experimental results}\label{sec:experimental_results}

\begin{figure}
  \includegraphics[width=\textwidth]{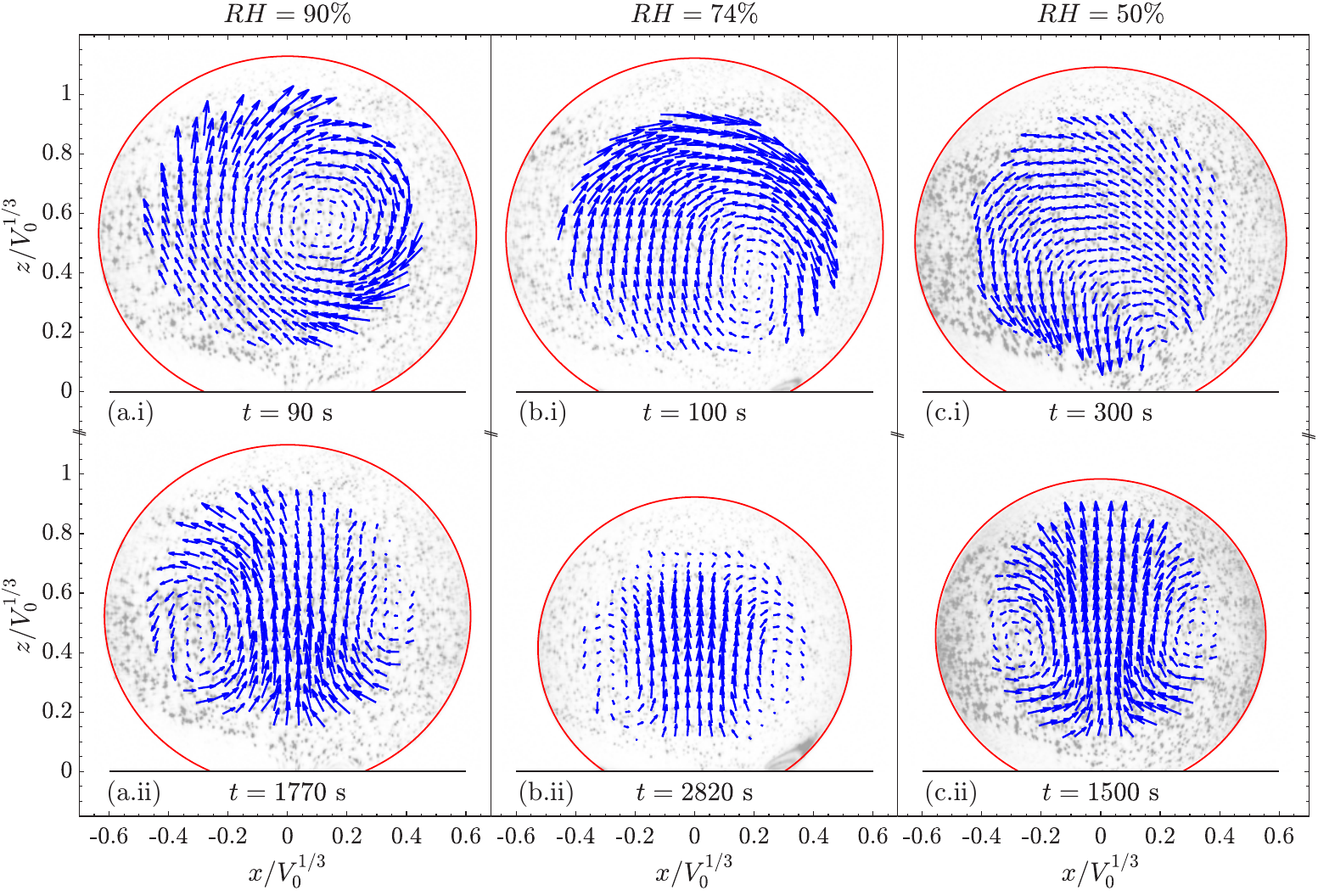}
  \caption{Experimental measurements of the flow in an evaporating drop. 
  Each column corresponds to a measurement at a different relative humidity. 
  \reviewerA{The plane $y=0$ is shown in the figure.}
  The evaporation speed increases from left to right. 
  Snapshots at two different times are shown. 
  Initially, the flow is asymmetric, with a single roll in a different direction for each experiment (upper row of figures). 
  After some time, the flow becomes axisymmetric in all experiments, with a flow from the apex towards the contact line, indicating dominant thermal buoyant forces (lower row of figures).
  }
  \label{fig:experiments}
\end{figure}

\reviewerC{Figures \ref{fig:experiments}(a-c) show 3 different experiments conducted at different relative humidity, namely \SI{90}{\percent}, \SI{74}{\percent}, and \SI{50}{\percent}, therefore having different evaporation rates.}
In these experiments, the flow starts as a single convection roll vortex within the droplet (upper row of figure \ref{fig:experiments}), to then eventually transition to axisymmetric flow pattern (lower row of figure \ref{fig:experiments}). The initial single roll flow pattern remains stable for several minutes (see supplementary movies 1-3), indicating that it is not merely a transient phenomenon.

\reviewerA{The direction of the observed axisymmetric interfacial velocity field in the late phase, from the apex towards the contact line, and upward-directed at the symmetry axis, suggests that the flow is driven by a thermal buoyant circulation (from here on \emph{axisymmetric Rayleigh flow}), and not a thermally-driven Marangoni flow (from here on \emph{axisymmetric Marangoni flow}).}

\reviewerA{However, such a flow pattern transition is not always observed, even though the experiments are run under identical experimental conditions. 
The flow transition illustrated in Figure \ref{fig:experiments} occurs in 40\% of the experiments done (a total of 50 experiments), and an additional 43\% of the experiments show either only a single roll flow pattern or only the  axisymmetric Rayleigh flow, with no observable transition. A minority of them (4\%) showed an inverted transition, from axisymmetric Rayleigh flow to single roll. 
Finally, a small but not negligible fraction of the experiments (13\%) showed a transition from an axisymmetric Marangoni flow, characterised by a downward-directed flow in the symmetry axis, into a axisymmetric Rayleigh flow, characterised by upward-directed flow. 
The variety of observed flow patterns reveal the high sensitivity of the transition due to the presence of contaminants in the environment.
Despite the diversity of flow patterns observed, the most frequently observed flow patterns are clearly dominated by thermally-driven convective flows, as those shown in Figure \ref{fig:experiments}. 
}

Such experimental results contrasts with the theoretical predictions \citep{hu2005analysis,saenz2015evaporation,diddens2017evaporating} that suggest that thermal Marangoni flows should be dominant, manifesting interfacial flows directed from the contact line towards the apex, and downward-directed flow in the symmetry axis.

This feature resembles the case of sessile droplets with lower contact angles, in which thermally-driven Marangoni flows are expected to develop flow patterns even stronger than the capillary flows. However, in practice, such thermally-driven flows are rather weak and only inconsistently observed \citep{hu2005analysis,gelderblom2022evaporation}.
The most accepted explanation of this surprising feature is that the presence of contaminants at the surface hinders the thermally-driven surface tension gradient at the interface. 

Similarly, we hypothesise that the presence of contaminants at large contact angles is the cause for the dominance of convective flows, in the same way that capillary flows dominate the evaporation-driven flows in low-contact-angle sessile droplets \citep{deegan1997capillary,gelderblom2022evaporation}. \reviewerA{The presence of contaminants also explains the large diversity of flow patterns observed in our experimental results.}
\reviewerC{In the experiments of figure \ref{fig:experiments}, the time of the transition from single roll to axisymmetric Rayleigh flow is significantly different. 
We theorise that this is mainly due to different amount of contaminants present.}
\reviewerC{Experimentally, we have not observed a strong dependence of the flow pattern with the relative humidity.}


In the remaining of the paper, we explore the influence of the presence of contaminants at the droplet's interface by making use of numerical methods, as it has been done in previous studies \citep{savino2002buoyancy, hu2005analysis,hu2006marangoni, girard2008effect, xu2007marangoni, van2022competition}. 

\section{Numerical models: governing equations}\label{sec:governing_equations}
\subsection{Transient model}\label{sec:transient_model}

We consider a simple model that captures the essential physics of the experiments. 
A similar model has been used by \citet{diddens_li_lohse_2021} for the prediction of the flow inside an evaporating binary glycerol-water droplet. In here, however, the temperature field inside a \textit{pure} water droplet is considered, rather than the compositional change of the species.

\begin{figure}
  \centerline{\includegraphics[width=1\textwidth]{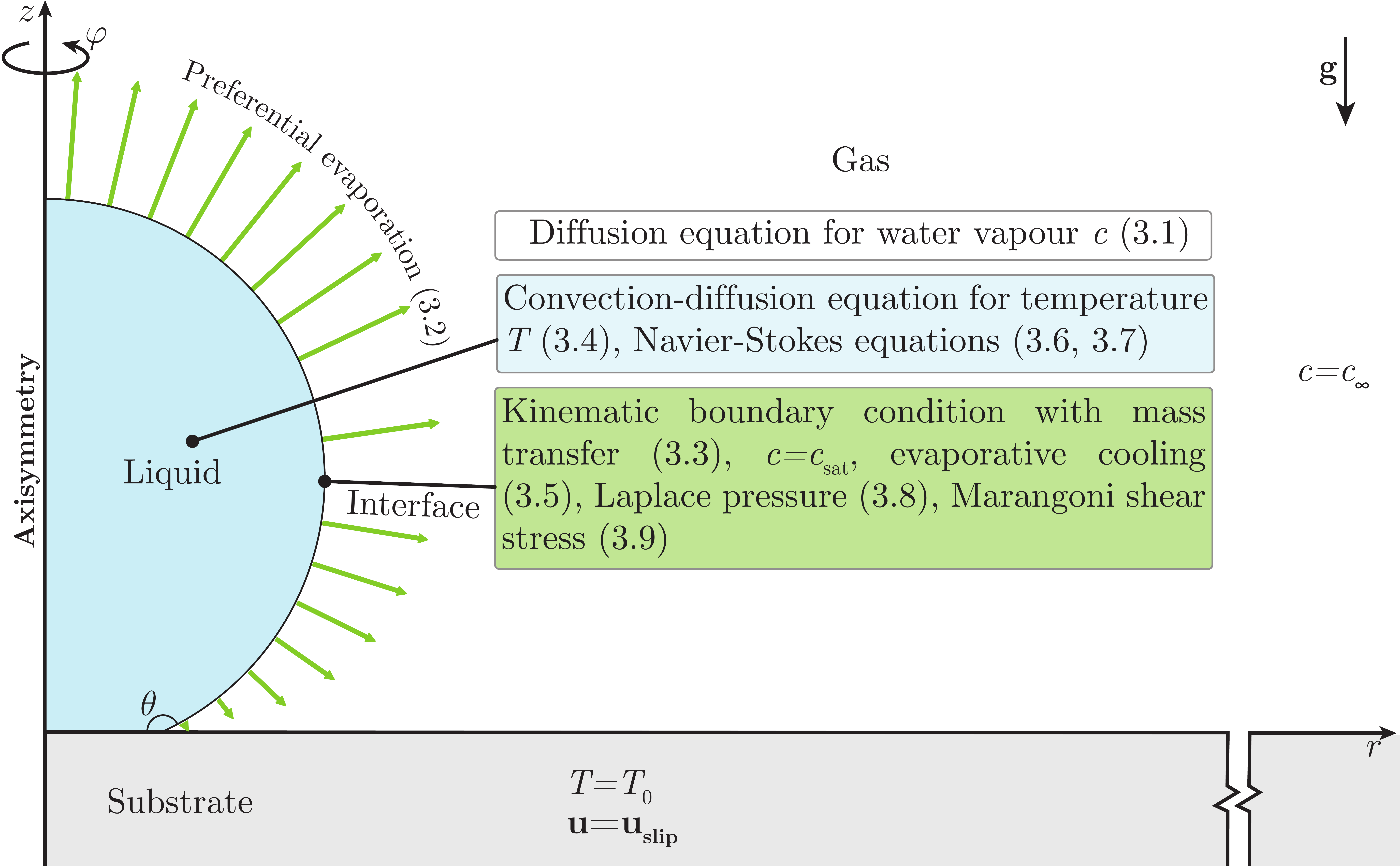}}
  \caption{Schematics of the model used in the numerical simulations.}
  \label{fig:schematics_model}
\end{figure}

A droplet of initial volume $V_0$, density $\rho$, viscosity $\mu$, thermal conductivity $k$ and specific heat capacity $c_p$ is deposited on a substrate, forming a contact angle $\theta$, see figure \ref{fig:schematics_model}. 
The substrate is approximated to be perfectly thermally conductive, such that its temperature is set to the ambient temperature $T_0$, which is far below the boiling point.
Considering finite thermal conductibility would result in lower thermal gradients within the droplet, weakening the thermal Marangoni effects.
However, our numerical tests with the experimental thermal conductibility of the substrate suggest that the interfacial velocity remains of the same order of magnitude as by considering a perfectly thermally conductive substrate, thereby justifying neglecting this effect.
The droplet is surrounded by air at $T_0$ and relative humidity $RH = c_\infty/c_\text{sat}$, where $c_\infty$ is the vapour concentration far from the droplet and $c_\text{sat}$ is the vapour saturation concentration at atmospheric pressure and constant $T_0$.
All liquid properties are temperature-dependent.

We consider the gas surrounding the water to be in a quiescent state.
The solutal and thermal Rayleigh numbers of the water vapour around the droplet are defined as follows \citep{dietrich2016role}: $\mathrm{Ra}_c^g = (\beta_c^g g V)/(\mu^g \kappa_c^g)$ and $\mathrm{Ra}_T^g = (\beta_T^g g V)(\mu^g \kappa_T^g)$, respectively. 
Here, $\beta_c^g = (\partial \rho^g / \partial c)/\rho^g$ is the solutal expansion coefficient, $\rho^g$ is the gas density, $c$ is the gas vapour concentration, $g$ is gravitational force, $\mu^g$ is the gas' dynamic viscosity, $\kappa_c^g$ is the solutal vapour's diffusivity, $\beta_T^g = (\partial \rho^g / \partial T)/\rho^g$ is the thermal expansion coefficient and $\kappa_c^g$ is the thermal vapour's diffusivity.
When calculating $\mathrm{Ra}_c$ and $\mathrm{Ra}_T$ for the gas and vapour around the droplet, we obtain $\mathrm{Ra}_s = 1.2$ and $\mathrm{Ra}_T = 0.5$, both much smaller than the critical Rayleigh value of 12 for convective effects to become relevant.
Since there are no external heat sources, the water vapour concentration $c$ can be considered to quasi-stationarily diffuse far from the droplet \eqref{eq:c_diffusion} and the evaporation rate $j$ to be given by diffusive flux at the liquid-gas interface \eqref{eq:evaporation_rate}\citep{deegan1997capillary,deegan2000contact,deegan2000pattern,popov2005evaporative}:
\
\begin{equation}
    \nabla^2 c = 0 \, ,
    \label{eq:c_diffusion}
\end{equation}
\begin{equation}
    j = - D^\text{vap} \bnabla c \bcdot \mathbf{n} \, ,
    \label{eq:evaporation_rate}
\end{equation}
\
\noindent where $D^\text{vap}$ is the vapour diffusion coefficient and $\mathbf{n}$ is the normal to the interface.
As the droplet evaporates, the droplet-gas interface moves according to the kinematic boundary condition \eqref{eq:kinematic_bc}:
\
\begin{equation}
    \rho (\mathbf{u} - \mathbf{u}_\text{i}) \bcdot \mathbf{n} = j
    \label{eq:kinematic_bc}
\end{equation}
Here, $\mathbf{u}$ is the velocity field and $\mathbf{u}_\text{i}$ is the velocity of the interface. 
On account of the latent heat of evaporation $\Lambda$, evaporative cooling occurs. 
Due to the presence of the isothermal substrate and the spatially non-uniform evaporation rate, temperature gradients along the interface are expected.
In the liquid phase, if convective motion is initiated through e.g. thermal Marangoni forces, it is expected to have significant effects on the droplet temperature profile due to the low thermal diffusivity $\kappa = k/(\rho c_p)$ of water (or any similar liquid), which is of order $\kappa \sim$ \SI{e-7}{\meter^2\per\second}.
Therefore, we consider a convection-diffusion equation for the temperature $T$ in the liquid phase \eqref{eq:heat_equation}:
\
\begin{equation}
    \rho c_p ( \partial_t T + \mathbf{u} \cdot \bnabla T ) = k \nabla^2 T  \, ,
    \label{eq:heat_equation}
\end{equation}
\
\noindent with a boundary condition for the evaporative cooling in a frame co-moving with the interface \eqref{eq:temperature_bc}:
\
\begin{equation}
    k \bnabla T \bcdot \mathbf{n} = - j \Lambda \, ,
    \label{eq:temperature_bc}
\end{equation}
\
\noindent and $T=T_0$ at the substrate. Thermal transport in the gas phase is disregarded due to its small contribution in the droplet's flow features.

We consider the z-axis to be the vertical axis pointing to the apex of the droplet, and gravity $g$ acts in the negative z-direction for a sessile droplet, a configuration to which this study is restricted.
The velocity field and pressure $p$ are governed by the Navier-Stokes equation: 
\
\begin{equation}  
  \rho ( \partial_t \mathbf{u}+ \mathbf{u} \bcdot \bnabla \mathbf{u} ) = - \bnabla p + \mu \nabla^2 \mathbf{u} + \rho \mathbf{g} \, .
  \label{eq:navier_stokes}
\end{equation}
\
\noindent where $\mathbf{g} = - g \mathbf{e}_z$ is the gravitational acceleration ($\mathbf{e}_z$ is the unit vector in the z-direction, positive upwards).
The mass conservation is ensured by the continuity equation:
\
\begin{equation}
  \partial_t \rho + \bnabla\bcdot (\rho\mathbf{u})=0 \, .
  \label{eq:continuity}
\end{equation}
\
Similarly to what is considered by \citet{diddens_li_lohse_2021}, we take a small slip-length to simulate the assumed pinned contact line in CR-mode \citep{picknett1977evaporation} to resolve the incompatibility of no-slip at the substrate and evaporation at the contact line.
As long as the slip-length is orders of magnitude smaller than the droplet size, the results are independent of the slip-length \citep{diddens_li_lohse_2021}, which is also confirmed by our simulations.

The liquid-gas interface must account for the Laplace pressure and Marangoni stresses, which are given by:
\
\begin{equation}
    \mathbf{n} \bcdot ( - p \mathbf{I} + \mu (\bnabla \mathbf{u} + (\bnabla \mathbf{u})^T) ) \bcdot \mathbf{n} = \sigma C \mathbf{n} \, ,
    \label{eq:laplace_pressure}
\end{equation}
\noindent and
\begin{equation}
    \mathbf{n} \bcdot (\mu (\bnabla \mathbf{u} + (\bnabla \mathbf{u})^T) ) \bcdot \mathbf{t} = \bnabla_t \sigma \bcdot \mathbf{t} \, .
    \label{eq:marangoni_stress}
\end{equation}
\
\noindent Here, $\mathbf{t}$ is the tangential vector to the interface, $\mathbf{I}$ is the identity matrix, $C$ is the curvature of the interface and $\bnabla_t$ is the surface derivative at the interface.

\subsection{Quasi-stationary model}\label{sec:quasi_stationary_model}

In a droplet evaporating at ambient conditions, the velocity $\mathbf{u}$ associated with the circulation within the droplet is typically much faster than the velocity of the interface $\mathbf{u}_\text{i}$ \citep{diddens2017evaporating}. 
Buoyancy and thermal Marangoni forces form a quasi-equilibrium, thereby justifying the quasi-stationary approximation.
The kinematic boundary condition \eqref{eq:kinematic_bc} is then reduced to $\mathbf{u} \cdot \mathbf{n} = 0$, which can be imposed numerically via a Lagrange multiplier field at the interface.
Due to its small size, the temperature variations within the droplet are typically a few percent of the absolute ambient temperature. 
It is then reasonable to consider an expansion of the temperature into $ T(\mathbf{x},t) = T_0 + \Delta T(\mathbf{x},t)$, i.e.\ the ambient temperature summed with the small local variations $\Delta T(\mathbf{x},t)$. 
Taking these approximations, it is logical to expand the density and surface tension into first-order Taylor series around the ambient temperature $T_0$:
\
\begin{equation}
    \rho(T) = \rho_0 + (T - T_0) \left. {\partial_T \rho} \right|_{T_0} \, ,
    \label{eq:rho_expansion}
\end{equation}
\begin{equation}
    \sigma(T) = \sigma_0 + (T - T_0) \left. {\partial_T \sigma} \right|_{T_0} \, ,
    \label{eq:sigma_expansion}
\end{equation}
\
\noindent where $\rho_0$ and $\sigma_0$ are the density and surface tension at $T_0$ and $\left. {\partial_T \rho} \right|_{T_0}$ and $\left. {\partial_T \sigma} \right|_{T_0}$ are the first order derivatives of the density and surface tension with respect to the temperature, evaluated at $T_0$.
We simplify the notation to $\partial_T \rho$ and $\partial_T \sigma$ in the following.
All other properties are considered to be constant and are evaluated at $T_0$, i.e.\ $\mu=\mu_0$, $k=k_0$, $c_p=c_{p0}$, $D^\text{vap}=D^\text{vap}_0$, $\Lambda = \Lambda_0$, $c_\text{sat}=c_\text{sat0}$, i.e.\ we assume the Oberbeck-Boussinesq-approximation for these properties.

We scale the vapour concentration such that $\tilde{c}=1$ at the droplet-gas interface and $\tilde{c}=0$ far away. 
For the temperature differences $T-T_0$, one should take into consideration that no temperature difference should be expected when no evaporation happens, i.e.\ when relative humidity is 100\% (or $c_\text{sat}=c_\infty$). 
We choose the factors in evaporative cooling \eqref{eq:temperature_bc} for the scaling of the temperature variation:
\
\begin{equation*}
    \begin{aligned}
        (c-c_\infty) &= (c_\text{sat}-c_\infty)\tilde{c} \, , \qquad\qquad& 
        (T-T_0) &= \dfrac{(c_\text{sat}-c_\infty) \Lambda_0 D^\text{vap}_0}{k_0} \tilde{T} \, .
    \end{aligned}      
\end{equation*}
\
We take the cubic root of the volume to nondimensionalise length scales.
The thermal diffusivity $\kappa_0 = k_0/(\rho_0 c_{p0})$ is taken to nondimensionalise time, since thermal effects dictate the flow.
We then have:
\
\begin{equation*}
    \begin{aligned}
        \mathbf{x} &= V^{1/3} \tilde{\mathbf{x}} \, , \qquad\qquad &
        t &= \frac{V^{2/3}}{\kappa_0} \tilde{t} \, , \qquad\qquad &
        \mathbf{u} &= \frac{\kappa_0}{V^{1/3}} \tilde{\mathbf{u}} \, . &
    \end{aligned}
\end{equation*}
\
A dimensionless mass transfer rate $\tilde{j}$ can then be defined as $\tilde{j} = V^{1/3}/(D^{vap}_0 (c_{sat}-c_\infty)) j = \bnabla \tilde{c} \bcdot \mathbf{n}$.
In nondimensional form, equations \eqref{eq:heat_equation} and \eqref{eq:temperature_bc} can be rewritten as:
\
\begin{equation}
  \partial_{\tilde{t}} \tilde{T} + \tilde{\mathbf{u}} \bcdot \tilde{\bnabla} \tilde{T} = \tilde{\nabla}^2 \tilde{T} \, , 
\end{equation} 
\begin{equation}
  \tilde{\bnabla} \tilde{T} \bcdot \mathbf{n} = - \tilde{j} \, .
\end{equation}
\
The Oberbeck-Boussinesq-approximation is taken for the Navier-Stokes equations since $(T-T_0) \partial_T \rho$ is small compared to $\rho_0$. 
The density appears as in equation \eqref{eq:rho_expansion} only in the gravity term $\rho \mathbf{g}$, while considered constant $\rho_0$ in the remaining terms.
Due to the small Reynolds number ($\sim 0.1$ at experimental conditions), \reviewerB{Stokes flow is assumed to be sufficient}. Equations \eqref{eq:navier_stokes} and \eqref{eq:continuity} can then be reduced to:
\
\begin{equation}
    \tilde{\bnabla} \bcdot \tilde{\mathbf{u}} = 0 \, ,
\end{equation}
\begin{equation}
    - \tilde{\bnabla} \tilde{p} + \tilde{\nabla}^2\mathbf{u} + \RAT \tilde{T} \mathbf{e}_z = 0\, ,
    \label{eq:navier_stokes_dimensionless}
\end{equation}
\
\noindent where:
\
\begin{equation*}
  \tilde{p} = \frac{V^{2/3}}{\kappa_0 \mu_0} (p - \rho_0 g z) \, 
\end{equation*}
\
\noindent
is the nondimensional pressure and:
\
\begin{equation*}
      \RAT = \frac{V g |\partial_{\tilde T} \rho|}{\kappa_0 \mu_0} \, .
\end{equation*}
\
\noindent
is the thermal Rayleigh number.

To remove the null space of the pressure field, we use a Lagrange multiplier which ensures that the average pressure $\tilde{p}$ in the liquid domain is zero.
Note that we derive $\rho$ with the nondimensional temperature $\tilde T$ to ensure that in the absence of evaporation, e.g.\ when $c_\text{sat}=c_\infty$, $\RAT=0$.
The nondimensional capillary \eqref{eq:laplace_pressure} and Marangoni \eqref{eq:marangoni_stress} forces are given by:
\
\begin{equation}
  \tilde{p} + \mathbf{n} \bcdot (\tilde{\bnabla} \tilde{\mathbf{u}} + (\tilde{\bnabla} \tilde{\mathbf{u}})^T) \bcdot \mathbf{n} = \frac{1}{Ca} (\tilde{C} + Bo \tilde{z}) \, ,
\end{equation}
\begin{equation}
  \mathbf{n} \bcdot (\tilde{\bnabla} \tilde{\mathbf{u}} + (\tilde{\bnabla} \tilde{\mathbf{u}})^T) \bcdot \mathbf{t} = - \MAT \tilde{\bnabla}_t \tilde{T} \cdot \mathbf{t} \, ,
  \label{eq:marangoni_stress_dimensionless}
\end{equation}
\
\noindent where the  capillary number $Ca$, the Bond number $Bo$ and the thermal Marangoni number $\MAT$ are given by:
\
\begin{equation*}
  \begin{aligned}
      Ca = \frac{\kappa_0 \mu_0}{V^{1/3} \sigma_0} \, , \qquad\qquad &
      Bo = \frac{\rho_0 g V^{2/3}}{\sigma_0} \, , \qquad\qquad &
      \MAT = \frac{V^{1/3} |\partial_{\tilde T} \sigma|}{\kappa_0 \mu_0} \, .
  \end{aligned}
\end{equation*}
\
In small droplets, the capillary number is small ($\sim$ \num{e-6} at the experimental conditions). 
\reviewerB{Since the Bond number is also small ($\sim 10^{-1}$ at the experimental conditions), the last term of equation \eqref{eq:laplace_pressure} is assumed to be irrelevant}, leading to an equilibrium spherical-cap shape with a constant contact angle \CA .

In our model, we have considered only the case of decreasing density and surface tension with increasing temperature, which is the case for water at ambient lab conditions.
This allows to consider only positive $\RAT$ and $\MAT$ values, following the sign convection in the definition of the system of equations.

\subsection{Procedure}\label{sec:procedure}
The system of equations of \S\ref{sec:transient_model} and \S\ref{sec:quasi_stationary_model} are solved using a finite element method on triangular elements with the package \software{pyoomph}\footnote{Available at: \url{https://pyoomph.github.io/}} \citep{diddens2024bifurcation}, which is based on \software{oomph-lib} \citep{heil2006oomph} and \software{GiNaC} \citep{bauer2002ginac}. 
The code was mutually validated by an analogous implementation in \software{ngsolve} \citep{schoberl1997netgen}.
The mesh motion in the transient model of \S\ref{sec:transient_model} is modelled by using an arbitrary Lagrangian-Eulerian (ALE) method.
Since the droplet geometry is assumed to be axially symmetric, a two-dimensional mesh is used in all simulations, even when analysing the azimuthal symmetry-breaking phenomenon (\S\ref{sec:phase_diagram_axisymmetry_breaking}).
This can be done by assuming that the axial symmetry is broken in a bifurcation, due to instabilities that act on a azimuthal wave number $m$. 
We employ the method described and validated by \citet{diddens2024bifurcation}, a linear stability analysis method that allows investigating symmetry-breaking instabilities.
Together with bifurcation tracking and pseudo-arclength continuation methods, this tool allows identifying the critical parameters for which the symmetry is broken.

The vapour diffusion equation \eqref{eq:c_diffusion} in the gas phase is discretised using first-order continuous Lagrangian elements. 
Its solution then provides the value of the evaporation rate $\tilde{j}$ at the interface. 
In order to spatially approximate the velocity-pressure pair in the Stokes equation, MINI-elements \citep{arnold1984stable} are used. 
The temperature field is approximated using first-order continuous Lagrangian elements.
The choices of the latter two spaces are made in order to avoid spurious oscillations in the tangential velocity at the droplet-gas interface.
Further details concerning the choice of field spaces are provided in Appendix \ref{ap:numerical_discretisation}.

\reviewerA{
In \S 4.2, we calculate the percentage of the total volume of liquid associated with the stream function $\psi$ that flows in the direction from the apex towards the contact line.
This percentage is denoted as $\psi^+$.
The value of $\psi^+$ provides insight into the dominant forces driving the flow. 
Specifically: if $\psi^+ > 90\%$, the flow is considered to be dominated by buoyancy forces; if $\psi^+ < 10\%$, the flow is considered to be dominated by thermocapillary forces.
By fixing $\MAT$, we can formally consider $\RAT$ as a Lagrange multiplier to obtain the unique $\RAT$ values which satisfies each $\psi^+$ constraint.
Subsequently, we employ pseudo-arclength continuation on to trace the curves that delineate the regime where there is competition between Marangoni and Rayleigh effects.
}

\section{Pure water droplet: numerical results}\label{sec:pure_water_droplet}

\subsection{At experimental conditions: analysis of axisymmetric flow direction}\label{sec:comparison_with_experiments}

We take the initial experimental conditions of figure \ref{fig:experiments}(c) as input for the numerical simulations to compare the results.
More specifically, we consider here: RH=50\%, $T_0=21^\circ$C, $V_0=\SI{5.6}{\micro\liter}$ and $\theta=150^\circ$. 
The transient model of \S\ref{sec:transient_model} is used here to describe the time-evolution of the evaporating droplet.
The Young-Laplace equation is solved to obtain the slightly gravity-deformed droplet shape at deposition time, used as the initial geometry condition for our simulations.

\begin{figure}
    \centerline{\includegraphics[width=1\textwidth]{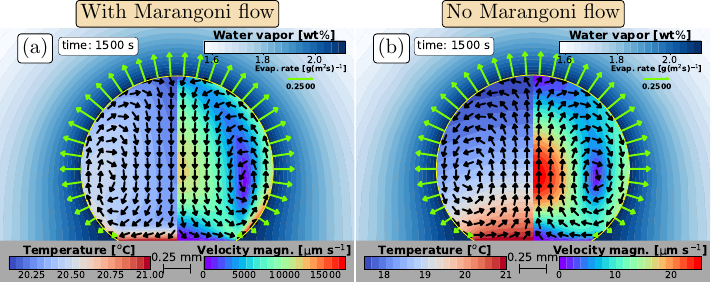}}
    \caption{Numerical results for \reviewerB{the temperature and velocity magnitude} fields at the initial experimental conditions of figure \ref{fig:experiments}(c), i.e. RH=50\%, $T_0=21^\circ$C, $V_0=\SI{5.6}{\micro\liter}$ and $\theta=150^\circ$, \SI{1500}{\second} after deposition on the substrate.
    Considering all effects generating the flow (a), the flow direction is axisymmetric, going from the rim to the apex, indicating dominant thermocapillary forces.
    If the surface tension is considered constant (b), the flow direction of the axisymmetric flow is the other way around, namely from the apex to the rim, indicating dominance of the buoyancy forces.
    }
    \label{fig:marangoni_vs_rayleigh}
\end{figure}

We analyse the flow direction obtained numerically in a time frame where the experiments present an axisymmetric flow from the apex towards the contact line, in particular \SI{1500}{\second} after deposition, as depicted in figure \ref{fig:experiments}(c.ii).
At this instant, the numerical calculations show a vortex in the droplet from the contact line towards the apex, indicating dominant thermocapillary forces, see figure \ref{fig:marangoni_vs_rayleigh}(a). 
The velocity magnitude is of the order of $\sim$\SI{10}{\milli\meter\per\second}, which is \reviewerB{significantly larger than} the experimental measurements, $\sim$\SI{1}{\micro\meter\per\second}.
One can theoretically consider a constant surface tension to disregard any Marangoni flow, see figure~\ref{fig:marangoni_vs_rayleigh}(b).
Here, despite the theoretical assumption, the flow direction agrees well with the experimental results, and the velocity magnitude is much closer to that of the experiments, $\sim$\SI{10}{\micro\meter\per\second}.
The discrepancy in flow direction and velocity magnitude between the numerical and experimental results is evident, indicating that there is an additional factor which significantly reduces the thermal Marangoni forces, resulting in a thermal-buoyancy-driven flow.

\subsection{Phase diagram of axisymmetric flow direction}\label{sec:phase_diagram}

\begin{figure}
  \centerline{\includegraphics[width=1.0\textwidth]{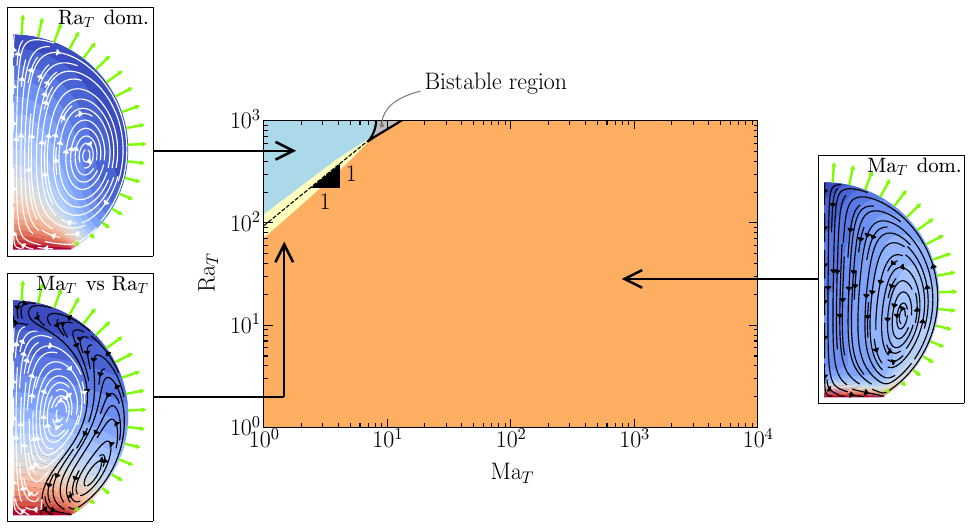}}
  \caption{$\MAT$-$\RAT$ phase diagram for flow direction for droplet of $\theta=$150$^\circ$.
  The grey area represents the bistable region, where multiple stable solutions coexist for the same $\MAT$ and $\RAT$ values.
  The blue region is dominated by thermal buoyancy, while the orange region is dominated by thermocapillary forces.
  In the yellow region, one observes both rolls driven by thermal Marangoni and by thermal buoyancy.
  The black dashed line represents the $\psi^+=50\%$ curve, where the competing vortices occupy equal volumes.}
  \label{fig:phase_diagram_MAG0}
\end{figure}

Hereinafter, the quasi-stationary model of \S\ref{sec:quasi_stationary_model} is used for the following calculations. 
We present a comprehensive analysis of the flow direction inside a droplet of $\theta=$150$^\circ$, in a $\MAT$-$\RAT$ phase diagram (figure \ref{fig:phase_diagram_MAG0}), assuming axial symmetry.
To ensure the accuracy of our findings, we first examine the stability of the stationary solution across the entire parameter space. 
\reviewerB{By calculating the eigenvalues of the linearised system of equations, we determine whether the obtained stationary solution is stable.}
From figure \ref{fig:phase_diagram_MAG0}, one can identify a small bistable region where multiple stable solutions coexist for the same $\MAT$ and $\RAT$ values. 
For simplicity, we exclude this region from our analysis, as indicated by the grey area in the phase diagram.
Then three remaining distinct regions can be identified in the phase space: the blue region, where buoyancy forces predominantly drive the flow, corresponding to $\MAT$ and $\RAT$ values such that $\psi^+ > 90\%$; the orange region, where thermocapillary forces primarily drive the flow, bounded by the curve $\psi^+ < 10\%$; and the yellow region, where two competing vortices are observed, defined by $10\% \le \psi^+ \le 90\%$. 

The black dashed line represents the $\psi^+ = 50\%$ curve, indicating where the competing vortices occupy equal volumes. 
Note the slope of 1 for this curve in this log-log phase space, which results from the linear combination of the effects of $\MAT$ and $\RAT$ at the onset of flow as $\MAT \rightarrow 0$ and $\RAT \rightarrow 0$.

The phase diagram (figure \ref{fig:phase_diagram_MAG0}) highlights a significant shift in flow direction with varying $\MAT$, evident in the thin transitional yellow region. 
\reviewerB{This shift can be attributed to the presence of warmer fluid in the bulk.
When a disturbance locally reduces the surface tension at the interface, a Marangoni flow transports fluid away from the perturbed region.
Incompressibility requires that this fluid is replenished from the bulk. 
Warmer fluid, with corresponding lower liquid-gas surface tension, is pushed towards the perturbed region, further increasing the Marangoni flow.
As a result, a self-enhancing thermal Marangoni is generated, leading to a rapid transition into a vortex from the contact line towards the apex as $\MAT$ increases.}
The phase diagram also reveals that, within the examined range of $\MAT$ and $\RAT$, Marangoni forces predominantly drive the flow for sufficiently large $\MAT$ (larger than $\sim 10$). 
Typically, $\RAT$ \reviewerB{is significantly smaller than} $\MAT$, rendering the blue region of the phase diagram unrealistic for water droplets. 
Thereby, the phase diagram also suggests that an additional factor that mitigates the influence of thermocapillary forces under the given experimental conditions must be considered to replicate the experimental observations.

\subsection{Axisymmetry-breaking}\label{sec:phase_diagram_axisymmetry_breaking}

\begin{figure}
  \centerline{\includegraphics[width=1.0\textwidth]{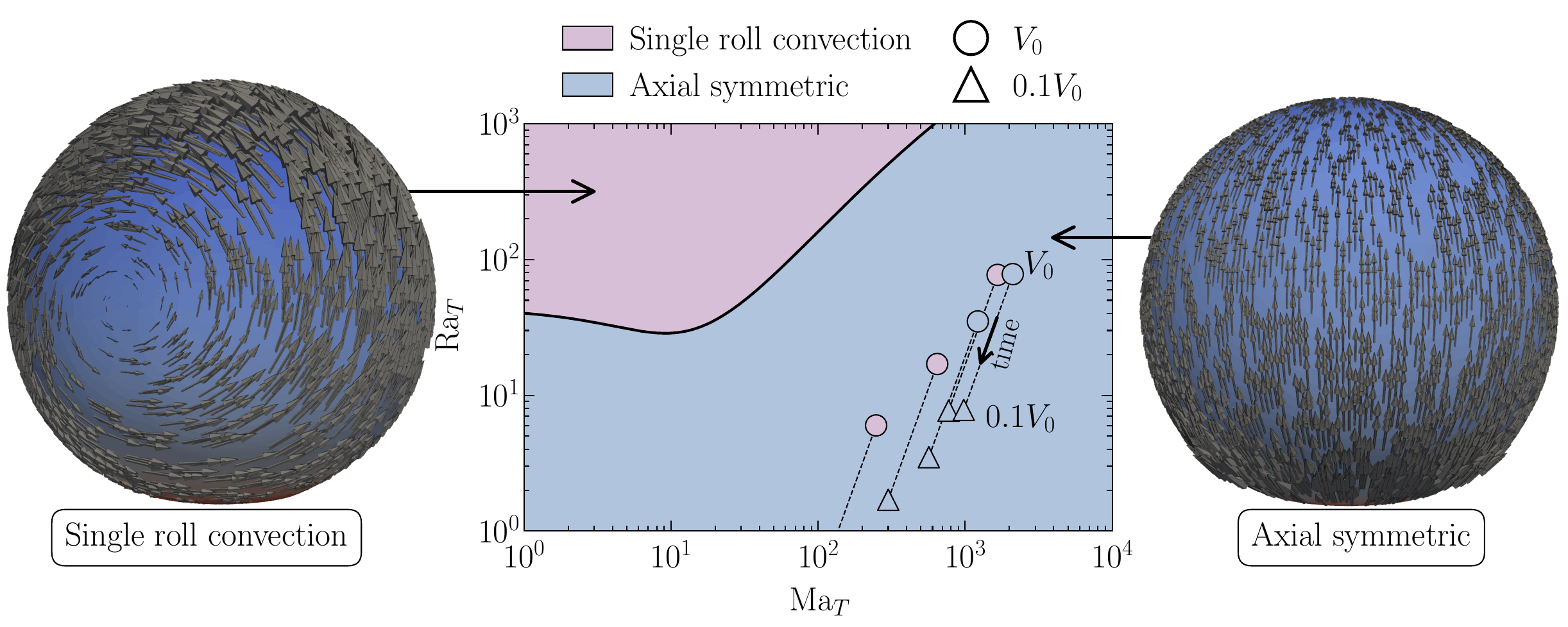}}
  \caption{Phase diagram of the axial symmetry-breaking phenomenon for a droplet of 150$^\circ$ contact angle.
  The black solid line represents the critical $\RAT^c$-$\MAT$ curve for which the flow loses stability of the axial symmetry solution (blue region) and can exhibit single-roll convection (\reviewerB{purple} region).
  The black dashed lines represent chosen experimental conditions, including those shown in figure \ref{fig:experiments}, initially (circular marker) and after $90\%$ of volume is evaporated (tringular marker).
  The experimental conditions (of Supplementary movies 1-5) are coloured according to the observed flow at the corresponding time instance, i.e.\ axisymmetric (blue) and single-roll convection (\reviewerB{purple}).
  Clearly, in 3 of the 5 cases here presented, the experimentally found single roll convection at the beginning of the evaporation process is in discrepancy with the theoretical expectation of this plot, reflecting that it is insufficient to only consider thermal effects as in section \ref{sec:pure_water_droplet}.
  In section \ref{sec:contaminants}, with considering also solutal Marangoni forces due to surfactants, this discrepancy is resolved.
  The right and left sides of the plot show the flow field for axial symmetry and for single-roll convection, respectively.}
  \label{fig:phase_diagram_axisymmetry_breaking}
\end{figure}

In the previous section, we assumed axial symmetry to determine the flow direction. 
Here, we investigate the stability of these stationary solutions under azimuthal symmetry. 
To do so, we use the method described in \S\ref{sec:procedure} to estimate the critical parameters at which azimuthal stability is broken and at which, simultaneously, the corresponding eigenfunction resembles a single-roll convection. 
We begin by setting $\MAT$=1 and iteratively adjusting $\RAT$ to find a value close to the critical $\RAT^c$ for axial symmetry-breaking at azimuthal wavenumber $m$=1. 
This is done by calculating the solutions of the eigenproblem at each iteration until the largest eigenvalue approaches zero, through a shift-invert Arnoldi's method \citep{arnoldi1951principle,saad2011numerical}.
Then, we activate the solver described by \citet{diddens2024bifurcation} to accurately find $\RAT^c$ at $\MAT$=1. 
Once $\RAT^c$ is determined, we obtain, through pseudo-arclength continuation, the critical $\RAT^c$-$\MAT$ curve, represented by the black solid line of figure \ref{fig:phase_diagram_axisymmetry_breaking}. 

The bifurcation curve represents the critical parameters for which the flow loses the axial symmetry (blue region) and can exhibit single-roll convection (\reviewerB{purple} region).
Note that the method used here (see \S\ref{sec:procedure}) does not provide the full nonlinear dynamics after symmetry breaking, which would require full three-dimensional simulations.
Bearing that in mind, the representations of possible flow fields in each region shown on the right and left sides of the figure are merely illustrative.
These flow representations depict axial symmetric and single-roll convection rolls, respectively, and they are obtained by revolving the two-dimensional flow field around the $z$-axis and assigning the total value of the perturbed velocity at each point, i.e.\ of $\mathbf{u}(r,\phi,z, t) = \mathbf{u}^0(r,z) + \epsilon \mathbf{u}^m(r,z)e^{i m \phi + \lambda_m t}+\text{c.c.}$, where $m=1$ and $r$, $\phi$ are the radial and azimuthal directions, $\epsilon$ is an attributed small amplitude of the perturbation, $\mathbf{u}^0$ is the axial symmetric part of the velocity field, $\mathbf{u}^m$ is the eigenfunction of the azimuthal perturbation and $\lambda_m$ is the corresponding eigenvalue.
Some chosen experimental conditions, including the ones shown in figure \ref{fig:experiments}, are depicted here for their respective initial conditions and after $90\%$ of the droplet volume has evaporated. 
We also performed azimuthal stability analysis for higher wavenumbers $m$, but no $\RAT^c$ were found within the considered phase space.

With an increase in $\MAT$, there is a corresponding rise in the critical $\RAT^c$, implying that thermal Marangoni effects contribute to the stabilisation of the flow in axial symmetry. 
This phenomenon can be linked to enhanced mixing within the droplet, a result of thermocapillary-induced convection. 
Interestingly, the experimental conditions all fall within the axisymmetric region, suggesting that the flow should maintain axial symmetry throughout the evaporation process. 
Contrary to this expectation, as discussed in \S\ref{sec:experiments}, the experimentally observed flows do not exhibit axial symmetry at the early stages. 
This discrepancy indicates, once again, the presence of an additional factor causing symmetry-breaking under the given experimental conditions, potentially due to a reduction in thermocapillary-induced mixing.

\section{Solutal Marangoni effects in contaminated water: numerical results}\label{sec:contaminants}

The presence of a tiny amount of contaminants, which can even be experimentally untraceable, can affect both the flow direction and the azimuthal symmetry within an evaporating droplet. 
We suggest that these contaminants are sufficient to function as surfactants and neutralise the surface tension gradients generated by thermal effects, thereby counteracting the impact of thermocapillary forces. 
Given the unknown characteristics of these contaminants, we model them as insoluble surfactants. 
These surfactants are then assumed to be present only at the droplet-gas interface, where they undergo diffusion along the interface and are carried by the interfacial velocity.
We start by introducing the transport equation for the surfactants' concentration $\Gamma$: 
\
\begin{equation}
  \partial_t \Gamma + \bnabla_t \bcdot (\mathbf{u}_\text{p} \Gamma) = D^\text{surf} \nabla_t^2 \Gamma \, .
  \label{eq:surfactant_transport}
\end{equation}
\
\noindent Here, $\mathbf{u}_\text{p}$ is the sum of the fluid velocity in tangential direction with the interface velocity in the normal direction, i.e.\ $\mathbf{u}_\text{p}= (\mathbf{I} - \mathbf{n}\mathbf{n})\mathbf{u} + (\mathbf{u}_\text{i} \bcdot \mathbf{n})\bcdot \mathbf{n}$.
Following a quasi-stationary approximation, $\mathbf{u}_\text{i} = 0$, such that $\mathbf{u}_\text{p} = (\mathbf{u}\bcdot \mathbf{t}) \mathbf{t}$.
To account for the unknown effect of contaminants on the surface tension, we assume a linear relationship between the decrease in surface tension and the surfactants' concentration, given by:
\
\begin{equation}
  \sigma(T, \Gamma) = \sigma_0 + (T - T_0) \left. \partial_T \sigma \right|_{\Gamma_0} + \Gamma \left. \partial_\Gamma \sigma \right|_{\Gamma_0} \, ,
  \label{eq:sigma_gamma}
\end{equation}
\
\noindent Here, $\left. \partial_\Gamma \sigma \right|_{\Gamma_0}$ is the first-order derivative of surface tension with respect to the surfactants' concentration, evaluated at the $\Gamma=0$.
Hereinafter, we simplify the notation to $\partial_\Gamma \sigma$.
It serves as a control parameter that determines the strength of the solutal Marangoni effect.

We define the dimensionless field $\tilde{\Gamma} = \Gamma / \Gamma_0$, where $\Gamma_0$ represents the average surfactants' concentration measured in \SI{}{\mol\per\meter^2}.
We impose a constraint to conserve the total concentration of surfactants on the droplet-gas interface $\partial \Omega$ = $\tau$, i.e.\ $\int_{\tau} \Gamma dS = \Gamma_0 A_0$, where $A_0$ is the initial surface area. 
The total surfactants' concentration is normalised such that, for a droplet of 150$^\circ$ contact angle, the dimensionless average concentration $\tilde{\Gamma}_0 =\int_{\tau} \Gamma dS / (A_0 \Gamma_0)$ is 1 at the beginning of the evaporation process (i.e. when $V=V_0$).

Taking the same scaling as in \S\ref{sec:quasi_stationary_model}, the dimensionless surfactant diffusion coefficient $D^\text{surf}$ is defined by the inverse of the Lewis number, $\operatorname{Le}^{-1} = D_0^\text{surf}/\kappa_0$. 
The diffusivity of insoluble surfactants is typically two orders of magnitude lower than the thermal diffusivity. 
To reduce the amount of parameters in the problem, we assume a fixed $D^\text{surf} =$\SI{e-9}{\meter^2\per\second}, i.e. $\operatorname{Le}\sim$\num{e2}.
Note that due to the unknown source of contaminants, it is difficult to estimate the exact value of $D^\text{surf}$.
In Appendix \ref{ap:lewis_number}, we briefly discuss the influence of the Lewis number on the flow direction.

The Marangoni stresses \eqref{eq:marangoni_stress} are affected by the change in surface tension due to surfactants. 
After scaling the governing equations appropriately, we can rewrite equation \eqref{eq:marangoni_stress_dimensionless} as:
\
\begin{equation}
  \mathbf{n} \bcdot (\tilde{\bnabla} \tilde{\mathbf{u}} + (\tilde{\bnabla} \tilde{\mathbf{u}})^T) \bcdot \mathbf{t} = \tilde{\bnabla}_t (- \MAT \tilde{T} - \MAG \tilde \Gamma ) \bcdot \mathbf{t} \, ,
  \label{eq:marangoni_stress_dimensionless_gamma}
\end{equation}
\
\noindent Here, the solutal Marangoni number $\MAG$ is a dimensionless quantity that represents the ratio between the reduction of surface tension gradient-induced advective transport due to surfactants and the viscous forces. 
It is defined as:
\
\begin{equation}
  \MAG = \frac{V^{1/3} |\partial_{\tilde \Gamma} \sigma|}{\kappa_0 \mu_0} \, .
  \label{eq:MAG}
\end{equation}
\
In order to give a physical understanding of the parameter, we consider water's physical properties and the initial volume of $V_0=\SI{5.6}{\micro\liter}$, corresponding to the initial experimental conditions of figure \ref{fig:experiments}(c).
Table \ref{tab:MAG} shows, in this context, the reduction of static surface tension compared to $\sigma_0$, and the corresponding $\MAG$ values.
\
\begin{table}
  \begin{center}
\def~{\hphantom{0}}
  \begin{tabular}{lccccccc}
      $\Gamma_0|\partial_\Gamma \sigma| / \sigma_0$ & 0.01 \% & 0.02 \% & 0.05 \% & 0.1 \% & 0.2 \% & 0.5 \% & 1 \% \\
      $\MAG$ & 91.77 & 183.5 & 458.9 & 917.7 & 1835 & 4588 & 9177 \\
  \end{tabular}
  \caption{$\MAG$ values corresponding to the initial reduction of surface tension $\Gamma_0 |\partial_\Gamma \sigma| / \sigma_0$ at the experimental conditions.}
  \label{tab:MAG}
  \end{center}
\end{table}
\
In the upcoming sections, we delve into the influence of surfactants on the flow direction under specific experimental conditions. 
We employ the quasi-stationary model (\S\ref{sec:quasi_stationary_model}) along with additional surfactants equations \eqref{eq:surfactant_transport}. 
\reviewerB{To ensure accurate and physically meaningful results, we have implemented a numerical scheme that prevents negative surfactants' concentrations, as detailed in Appendix~\S\ref{ap:numerical_discretisation}}. 
Later on, we consolidate our findings into a comprehensive $\MAT$-$\RAT$ phase diagram for fixed $\MAG$, for the flow direction assuming axial symmetry, and for the stability of the azimuthal symmetry.

\subsection{Flow reversal at experimental conditions} \label{sec:flow_reversal_experiment}

\begin{figure}
    \centerline{\includegraphics[width=\textwidth]{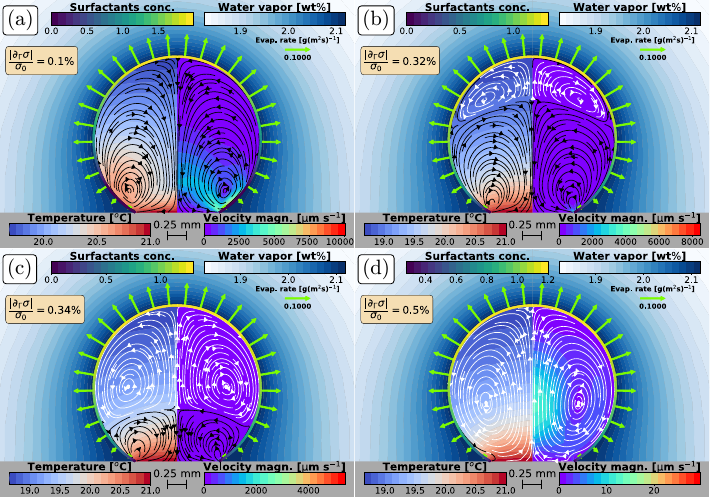}}
    \caption{Influence of surfactants on the flow direction. The values of $0.1\%$ (a), $0.32\%$ (b), $0.34\%$ (c), and $0.5\%$ (d) are considered for the reduction of static surface tension $| \partial_\Gamma \sigma | / \sigma_0$. 
    The flow direction is from the contact line towards the apex in (a); a competing vortex in the opposite direction is observed in (b) due to the increasing influence of the surfactants; rapid growth of the thermal Rayleigh vortex (c); the flow is completely in the thermal Rayleigh direction in (d), with a velocity magnitude comparable to the experimental findings.}
    \label{fig:surfactants_effects_exp}
\end{figure}

Using the same experimental initial conditions as in \S\ref{sec:comparison_with_experiments}, we investigate the influence of surfactants on the flow direction.
For a slight average reduction of 0.1$\%$ of the static surface tension $\sigma_0$ due to the action of surfactants, the flow follows the expected thermal Marangoni direction, from the contact line towards the apex, figure \ref{fig:surfactants_effects_exp}(a). 
However, if the reduction of $\sigma_0$ is 0.32$\%$, a competing vortex in the opposite direction emerges, figure \ref{fig:surfactants_effects_exp}(b). 
This opposing vortex is much larger if the surface tension reduction is 0.34$\%$, figure \ref{fig:surfactants_effects_exp}(c). 
Finally, when the reduction reaches 0.5$\%$, the flow completely reverses, moving from the apex towards the contact line, figure \ref{fig:surfactants_effects_exp}(d).
The findings from the study by \citet{van2022competition}, which do not account for buoyancy-driven convection, indicate that a reduction of approximately 0.8\% in surface tension results in a substantial decrease of 99\% in interface velocity. 
It is noteworthy that these results are consistent with our own findings, despite the omission of buoyancy effects in their paper.

The velocity magnitude in figure \ref{fig:surfactants_effects_exp}(d) is comparable to both the experimental results of figure \ref{fig:experiments}(b) and the hypothetical pure thermal Rayleigh flow depicted in figure \ref{fig:experiments}(b). 
These results demonstrate that even a tiny amount of contaminants can cause a significant change in the flow direction. 
It is also evident that for a larger reduction of $\sigma_0$, the distribution of the concentration of surfactants at the interface is more uniform. 
This is expected, as surfactants effectively reduce surface tension, resulting in weaker advective forces associated with thermal Marangoni flow. 
Consequently, surfactants diffuse along the interface, rather than concentrating at the apex of the droplet.

\subsection{Phase diagram of axisymmetric flow direction} \label{sec:flow_direction_phase_diagram}

\begin{figure}
  \centerline{\includegraphics[width=1\textwidth]{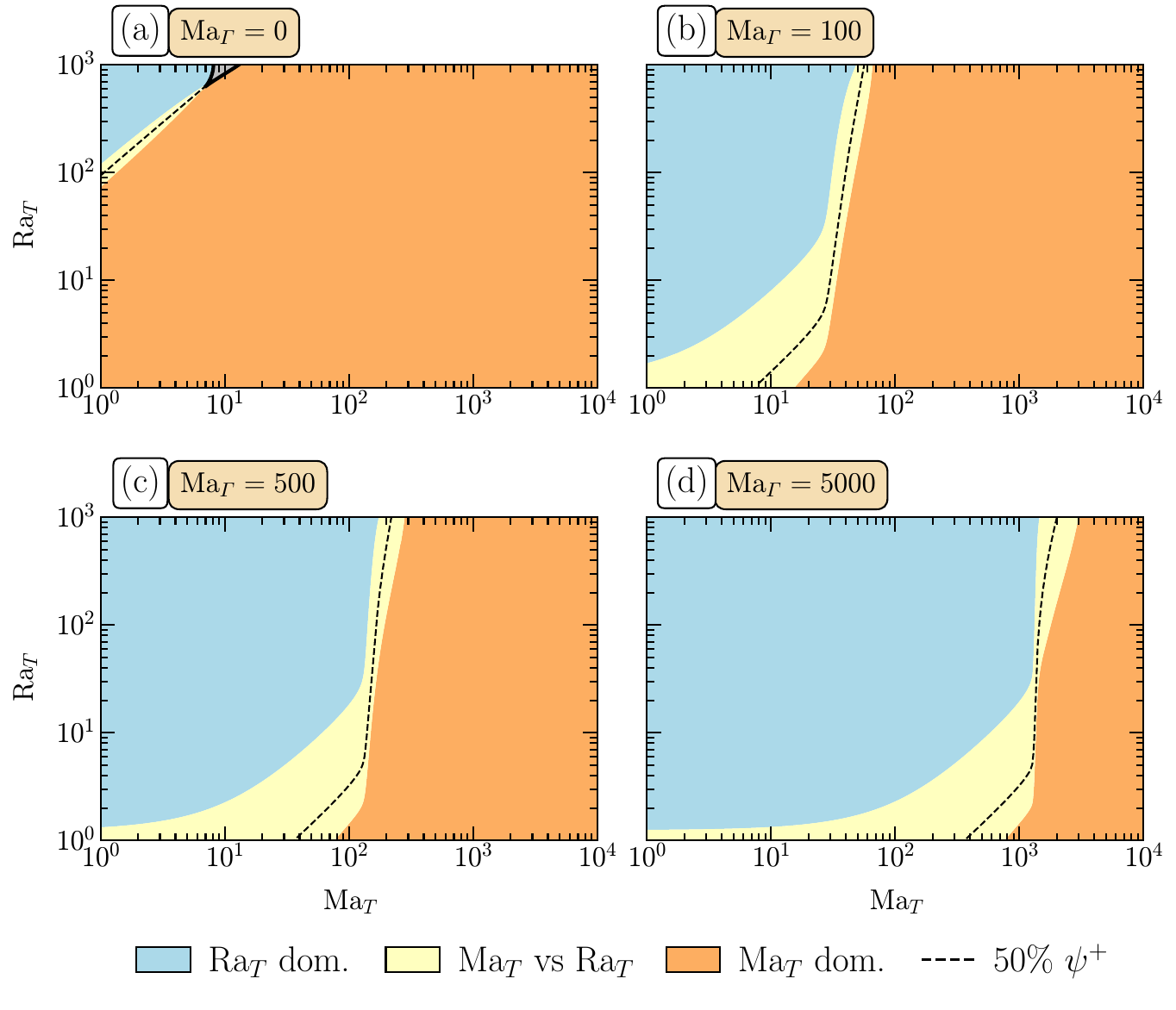}}
  \caption{Flow direction phase diagram for a droplet with a contact angle $\theta = 150^\circ$, considering the influence of surfactants at the interface, for $\MAG=0$ (a), $\MAG=100$ (b), $500$ (c) and $5000$ (d).
  The blue area corresponds to $\RAT$ dominated flows, the orange to $\MAT$ dominated flows, and the yellow to a combination of both.
  The black dashed line corresponds to 50$\%$ in volume of flow in each direction, i.e.\ $\psi^+=50\%$.}
  \label{fig:phase_diagram_largerMAG}
\end{figure}

When surfactants are taken into account, the phase diagram undergoes significant changes, see figure \ref{fig:phase_diagram_largerMAG}. 
Using the same colour code as in figure~\ref{fig:phase_diagram_MAG0}, we consider the $\MAG$ values: 0 (a), 100 (b), 500 (c), and 5000 (d).
Even a slight $\sigma_0$ reduction of approximately 0.01$\%$, corresponding to $\MAG = 100$ at the conditions specified for table \ref{tab:MAG}, leads to a substantial portion of the phase diagram being covered by the blue area, indicating flow in the apex towards the contact line. 
Remarkably, for $\MAG = 5000$, within the considered ranges of $\RAT$ and $\MAT$, the flow is almost always in the same direction as if driven by thermal buoyancy, i.e.\ moving from the apex towards the contact line. 
This corresponds to a mere $\sigma_0$ reduction of approximately 0.5$\%$.
Supplementary movie 6 illustrates the change in flow direction when travelling through the different regions in the phase diagram at $\MAG=100$.

The influence of the contact angle on the direction of the flow is discussed in Appendix \ref{sec:contact_angle_flow_direction_phase_diagram}.

\subsubsection{Change of slope in the $\psi^+=50\%$ curve} \label{sec:zero_surfactants}

Figure \ref{fig:phase_diagram_largerMAG} displays a change in the slope of the $\psi^+=50\%$ curve as $\MAT$ increases in the presence of surfactants. 
Notably, the quasi-stationary solutions along this curve feature two competing vortices — one driven by buoyancy and the other by Marangoni forces — that occupy equal volumes within the droplet.
We identify two distinct regimes of the curve corresponding to different slopes, for low and for high $\MAT$ values. 

\begin{figure}
\centerline{\includegraphics[width=1.0\textwidth]{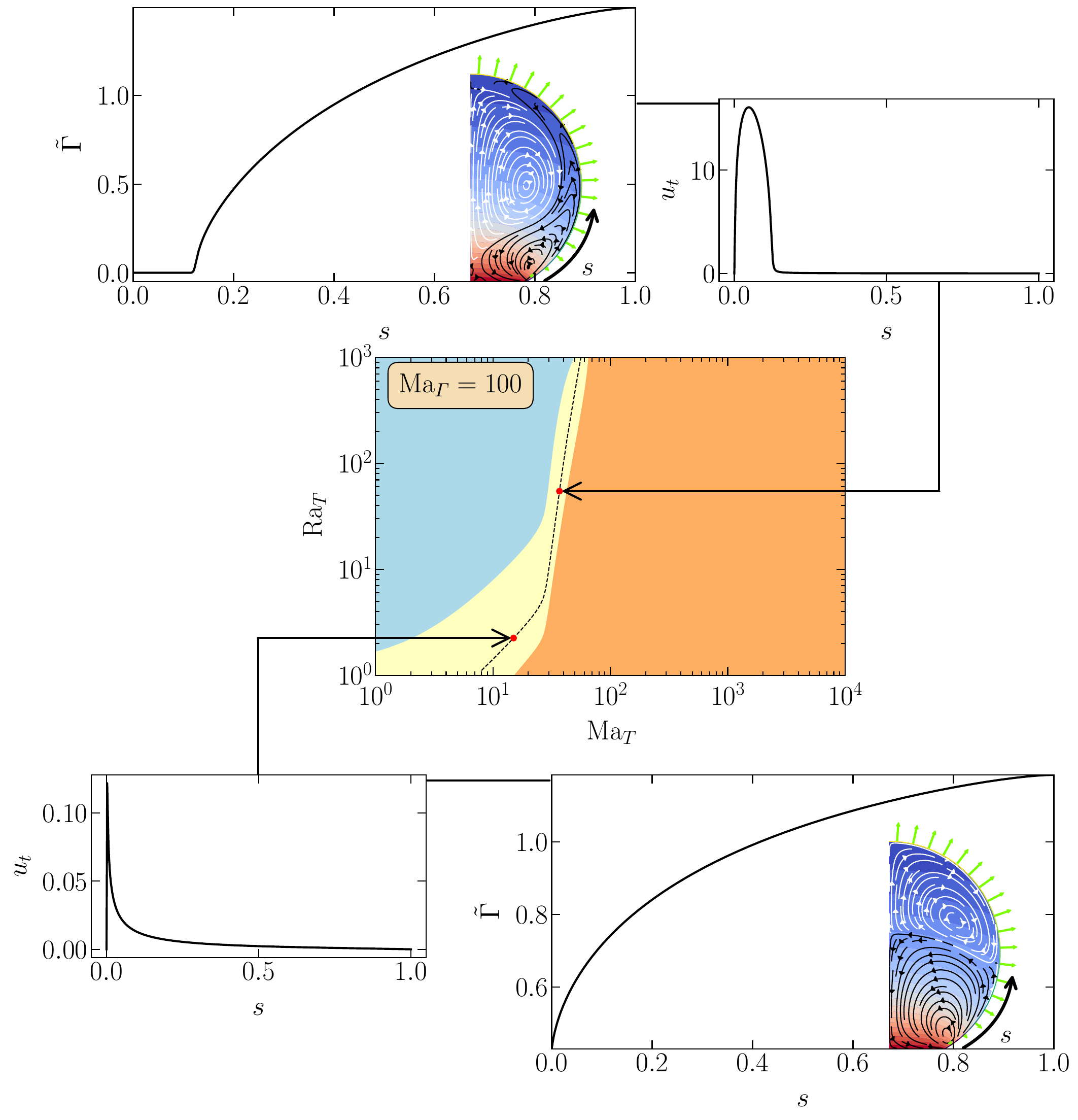}}
\caption{Surfactants' concentration along the droplet-gas interface according to the normalised droplet-gas arclength $s$, for $\MAG=100$, $\theta=150^\circ$ and two $\MAT$, one below the threshold for which the flow reversal curve changes slope (bottom) and one above (top).}
  \label{fig:surfactants_concentration}
\end{figure}

To explore the impact of surfactants on this phenomenon, figure \ref{fig:surfactants_concentration} examines a specific case from figure \ref{fig:phase_diagram_largerMAG} with $\MAG=100$. 
The surfactants' concentration along the droplet-gas interface is depicted for both the lower (bottom) and the higher (top) regimes of the curve. 
Additionally, the figure illustrates the vortices within the droplet for each corresponding $\MAT$ value and the tangential velocity $u_t$ at the droplet-gas interface, plotted against the normalised arclength coordinate $s$, where $s=0$ is at the contact line and $s=1$ is at the apex.

For low $\MAT$, the slope of the curve is approximately 1. 
In this regime, the vortex from the contact line towards the apex is primarily located near the droplet-gas interface, as shown at the bottom of figure \ref{fig:surfactants_concentration}. 
Thermocapillary forces counteract the effect of surfactants, resulting in a weak thermal Marangoni flow close to the interface, as depicted in the respective $u_t$-$s$ plot.

As $\MAT$ increases, surfactants are advected and compressed towards the droplet apex. 
In the high $\MAT$ regime (top part of figure \ref{fig:surfactants_concentration}), the surfactants' concentration at the contact line reaches zero, as shown in the respective $\tilde{\Gamma}$-$s$ plot for small $s$. 
Consequently, with no counteracting surfactant effect near the contact line, a strong thermal Marangoni flow emerges in that region, as indicated by the large $u_t$ in the top $u_t$-$s$ plot for small $s$. 
This flow is strong enough to reverse the flow direction not only at the interface but also near the axis, as illustrated by the vortex representations.

The change in the slope of the $\psi^+=50\%$ curve is directly caused by the surfactant concentration at the contact line dropping to zero at high $\MAT$ values. 
In the absence of surfactants near the contact line, a strong thermal Marangoni flow develops, easily overpowering the thermal Rayleigh forces. 
Supplementary movie 7 provides additional visual support for this explanation.

\subsection{Axisymmetry-breaking}\label{sec:axisymmetry_breaking}

\begin{figure}
  \centerline{\includegraphics[width=1\textwidth]{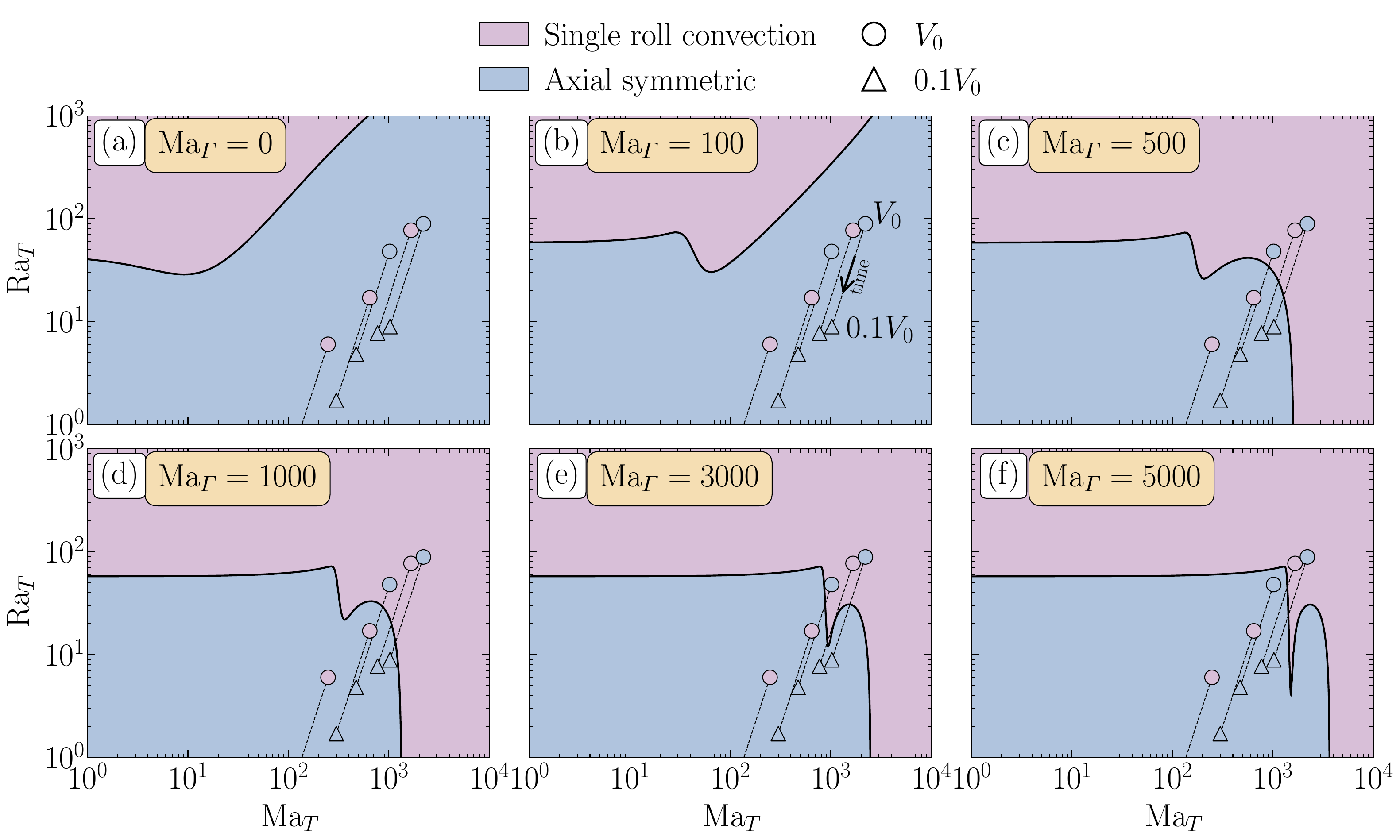}}
  \caption{Stability analysis at azimuthal wavenumber $m=1$ for $\MAG$=0 (a), $\MAG$=100 (b), $\MAG$=500 (c), $\MAG$=1000 (d), $\MAG$=3000 (e) and $\MAG$=5000 (f) of a droplet of 150$^\circ$ contact angle in a $\RAT$-$\MAT$ phase diagram. 
  The black solid line represents the critical $\RAT^c$-$\MAT$ curve for which the flow goes from axial symmetry (blue bottom part) to single roll convection (purple top part).
  The black dashed lines represent chosen experimental conditions (of Supplementary movies 1-5), including those shown in figure \ref{fig:experiments}.}
  \label{fig:axisymmetry_breaking}
\end{figure}

When surfactants are present ($\MAG>0$), the azimuthal symmetry of the flow is broken over a wider range of $\RAT$ and $\MAT$ than for the $\MAG=0$ case. 
Figure \ref{fig:axisymmetry_breaking} presents the results of the azimuthal stability analysis for various values of $\MAG$.

Interestingly, our calculations show that, for certain values of $\MAG$, the flow can initially exhibit asymmetric single-roll convection and, as evaporation progresses (or as the droplet volume decreases), the flow can eventually stabilise and regain axial symmetry.
In figure \ref{fig:axisymmetry_breaking}(c-f), we show one single experiment that displays this behaviour while presenting quantitative agreement between experiments and numerical simulations.
Nevertheless, the numerical simulations do not accurately capture the critical parameters for the transition from single-roll convection to axial symmetry across all experimental conditions.
This discrepancy may be due to simplifications in the numerical model, the absence of detailed information about the surfactants' properties, or sporadic experimental events such as contact line pinning.
Although our simplified model does not allow for precise quantitative comparison with the experiments, it does provide qualitative insight into the influence of contaminants on the stability of the axisymmetric quasi-stationary solutions.
For instance, when considering $\MAG=100$ and sufficiently large $\MAT$, as $\MAT$ increases, the critical $\RAT^c$ also rises, indicating that thermal Marangoni-induced circulation stabilises the flow and restores axial symmetry.
For $\MAG \gtrapprox 500$, the bifurcation curve becomes independent of $\MAT$ in the limit of low $\MAT$.
However, for larger $\MAT$, the critical $\RAT^c$ decreases with increasing $\MAT$, suggesting that the combined effects of solutal and thermal Marangoni forces destabilise the flow and break axial symmetry.

In the next section, we further explore the influence of the contact angle on the azimuthal symmetry-breaking phenomenon.
To simplify the problem, we isolate the thermal Rayleigh effects ($\MAT=\MAG=0$) and focus on examining how the large contact angle affects azimuthal symmetry-breaking.

\subsection{Contact angle influence in axisymmetry-breaking: pure Rayleigh flow} \label{sec:axisymmetry_breaking_theoretical}

\begin{figure}
    \centerline{\includegraphics[width=1.0\textwidth]{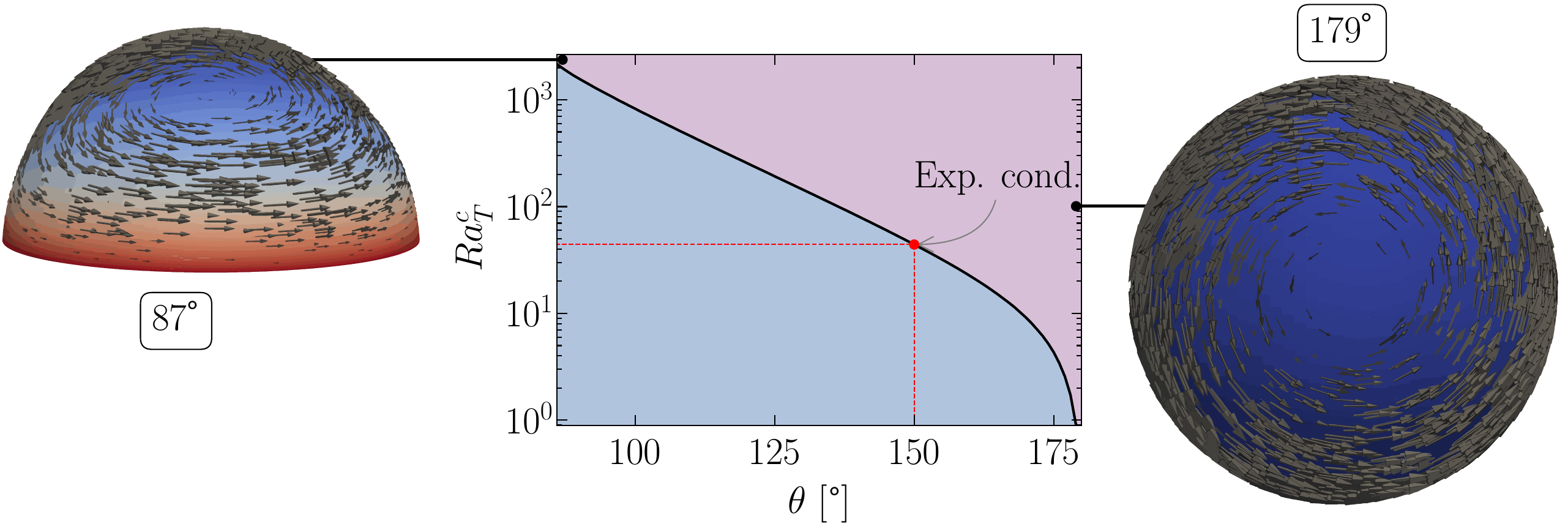}}
    \caption{Critical $\RAT^c$ for different contact angles \CA\ for the $m=1$ instability to happen, for $\MAG=\MAT=0$. 
    The black solid line represents the $\RAT^c$-$\MAT$ curve for which the flow goes from axial symmetry (blue bottom part) to single roll convection (purple top part).}
    \label{fig:axisymmetry_breaking_diff_CA}
\end{figure}

We present $\RAT^c$ for a range of \CA, from $85^\circ$ to $179^\circ$, figure \ref{fig:axisymmetry_breaking_diff_CA}, considering $\MAT=\MAG=0$. 
At the left of the figure, the convective roll flow field at $\theta=87^\circ$ is shown and, at the right, the flow field at $\theta=179^\circ$.
Obviously - and as the flow field representations reflects - for droplets of fixed volume, the contact radius will be much larger if the contact angle is smaller.
The contact area with the substrate, which has fixed temperature and is perfectly conducting,is much smaller for larger \CA, which will result in a larger temperature gradients within the droplet.
As $\theta \rightarrow 180^\circ$, the contact radius reduces to zero, approaching a heat point source. 
At the same time, an increasing contact angle reduces the size of the no-slip boundary condition at the substrate and thereby allowing for single roll convection, which is almost free from viscous forces in the limit $\theta \rightarrow 180^\circ$.

As the contact angle increases, $\RAT^c$ decreases, indicating that the large contact angle contributes to the azimuthal instability of the flow, similar to the onset of Rayleigh-B\'enard convection in a cylindrical container with aspect ratio of 1. 
To further support this hypothesis, we considered a theoretical scenario with no thermal buoyancy effects ($\RAT=\MAG=0$), only thermal Marangoni effects. 
The resulting solution of the azimuthal eigenvalue problem show that flow is always stable in the azimuthal direction, independently of $\MAT$.
Surfactants thus play a key role in the complex large $\MAT$ azimuthal symmetry-breaking phenomenon depicted in figure \ref{fig:axisymmetry_breaking}(c-f), as concluded from the results of this section.

For shallow droplets (small contact angle), thermal Marangoni instabilities are expected to induce highly nonaxisymmetric flow patterns \citep{babor2023linear}, also known as hydrothermal waves \citep{Sefiane2008}, which usually appear on heated substrates. Solutal Marangoni flow can exhibit even more asymmetric and even chaotic flow, as e.g. observed in evaporating ethanol-water droplets \citep{diddens2017evaporating}.
Since these phenomena are not observed in water droplets at ambient conditions, they are beyond the scope of this article.

\section{Outlook}\label{sec:conclusions}

Evaporating pure water droplets sitting on thermally conductive substrates exhibit, due to evaporative cooling, a temperature gradient within the droplet and along its droplet-gas interface.
As a result, an interplay between buoyant and thermocapillary forces are expected.
Previous theoretical and numerical predictions \citep{hu2005analysis,saenz2015evaporation,diddens2017evaporating}, as well as our simulations show that thermocapillary forces are dominant for small droplets, resulting in an interfacial thermally-driven Marangoni flow directed from the contact line towards the apex.
However, previous experimental observations \citep{buffone2004investigation,marin2016surfactant,diddens2017evaporating,rossi2019interfacial}, and our own measurements shown in \S\ref{sec:experiments}, reveal that the thermally-driven Marangoni flow is hindered by the presence of surfactants (when intentionally added) or contaminants (when its presence is unintentional).
In the case at hand, we show that thermally-driven Marangoni flows are rarely seen, and when present, their strength is orders of magnitude lower than expected.

A detailed transient model for a clean, pure water droplet was initially used to perform numerical simulations under the same conditions as in the experiments. 
The numerical results showed a significant discrepancy when compared to the experimental data. 
To address this, a small amount of insoluble surfactants was introduced in the axisymmetric numerical simulations. 
This adjustment led to a shift in the flow direction, aligning it with the thermal Rayleigh flow observed in most experiments, even when using a simplified quasi-stationary model.
Additionally, the quasi-stationary approach was employed to generate $\MAT$-$\RAT$ phase diagrams for the flow direction under the assumption of axisymmetry, and to analyse the azimuthal stability at an azimuthal wave number $m = 1$, for both pure and contaminated water cases.

At specified initial experimental conditions, as those shown in figure \ref{fig:experiments}(c), a mere 0.5$\%$ reduction in the static surface tension was sufficient to reverse the flow direction, agreeing remarkably well with the results of \citet{van2022competition}, despite their work not considering buoyancy-driven convection.
When surfactants are considered, the phase diagram for the flow direction undergoes significant changes, even for small $\MAG$.
This indicates that a larger set of parameters within the considered phase space will show a vortex from the apex towards the contact line.
Within this phase space, as $\MAT$ rises, the accumulation of surfactants at the apex increases up to a point where no surfactants are present close to the contact line.
A strong thermal Marangoni flow emerges in this region, rapidly reversing the flow upon a slight increase in $\MAT$.
For smaller \CA, the thermal profile within the droplet is more uniform due to the larger contact radius and smaller apex height. 
When surfactants are considered, a larger $\MAT$ is required for the droplet to present a vortex from the contact line towards the apex, as long as $\RAT$ is sufficiently big.
The onset of $m=1$ asymmetry is significantly affected by the presence of surfactants. 
For certain $\MAG$ values, the flow can initially show asymmetric single-roll convection, but as evaporation progresses, it may stabilise and regain axial symmetry. 
This numerically found behaviour matches that of the majority of our experiments, though the numerical simulations do not precisely capture the critical parameters for the transition to axial symmetry in all experiments.
Despite the model's simplicity , it provides an unique insight into the influence of contaminants on the stability of axisymmetric solutions. 
For $\MAG=100$ and large $\MAT$, the critical $\RAT^c$ increases, suggesting that thermal Marangoni circulation stabilises the flow and restores symmetry. 
For $\MAG \gtrapprox 500$, the bifurcation curve becomes independent of $\MAT$ at lower values. 
This implies that, in a droplet with a contact angle of $150^\circ$, the flow asymmetry is closely linked to the $\RAT$ number. 
In ideal scenarios in the absence of contamination or surfactants, our analysis predicts a single convective roll at lower $Ra_T^c$ for larger contact angles, indicating that the large contact angle contributes to azimuthal instability. 
However, for larger $\MAT$ and $\MAG \gtrapprox 500$, the critical $\RAT^c$ decreases, showing that combined solutal and thermal Marangoni forces destabilise the flow and break symmetry.

Contaminants thus induce such a significant change in the flow features.
The source of the contamination acting as surfactants is yet unknown, but can be likely found in both the original particle solutions (unavoidable to certain degree in the production process) and from the environment (organic material in the room's air).

The propensity of water-air interfaces to become polluted has been reported in multiple systems in the past, from rising bubbles \citep{squires2020surfactant} to liquid bridges \citep{montanero2016ponce}, and its prevention has always encountered only minor success. 
The impact of contamination from the particle solution can be mitigated by carefully washing the particle solution in multiple iterations. The results however show a even larger variety of flow patterns (including fading thermocapillary flows) than those observed with unwashed solutions, revealing a higher sensitivity of the interface to contamination from the environment, and in agreement with the results of   \citet{kazemi2021marangoni}. The thermocapillary flows observed under those conditions always quickly evolved into convective buoyant flow patterns during the evaporation process, denoting a likely increase in contamination over time in those conditions. Looking at the experimental results, the degree of contamination seemed more consistent for unwashed solutions. Therefore, we decided to only show those experimental results in this work. The issue with different degrees of contamination will be a topic for a future work. 

Although the present work, as well as those of \citet{van2022competition} and \citet{hu2005analysis}, provide a reasonable explanation for the absence of thermal Marangoni flow in evaporating water droplets, the presence of surfactants in multi-component droplets is still unexplored. 
The presence of contaminants in glycerol-water mixtures have been attributed as the cause for the formation of Marangoni rings upon evaporation \citep{thayyil2022evaporation}, but to the best of our knowledge, the influence of surfactants on the flow patterns in such mixtures has not yet been quantitatively investigated.

\section*{Acknowledgements}
This work was supported by an Industrial Partnership Programme of the Netherlands Organisation for Scientific Research (NWO) \& High Tech Systems and Materials (HTSM), 
co-financed by Canon Production Printing Netherlands B.V., University of Twente, and Eindhoven University of Technology. 
The authors would like to give a special acknowledgment to Carola Seyfert, whose sharp eye detected this phenomenon for the first time and her careful experiments inspired this paper. 
We are also thankful to Yongchen Jiang (exchange fellow USTC) and S\'ebastien Galas (exchange ENS Paris Saclay), who obtained the shown particular experimental data set during their internships in our labs.

\appendix

\section{Spatial discretisation}\label{ap:numerical_discretisation}

Both the droplet domain $\Omega$ and the gas domain $\Omega^\text{gas}$ are partitioned into a triangular mesh composed of elements $\mathcal{T}_h$ and $\mathcal{T}^\text{gas}_h$.
The $\Gamma$-field is only present at the two-dimensional droplet-gas interface, $\tau = \partial \Omega$.
When considering an axial symmetric coordinate system, $\Gamma$ is only present on a one-dimensional curved line, therefore embedded in a two-dimensional space.
We denote the set of 1D-elements composing the interface as $\mathcal{I}_h$, and denote by $\mathcal{F}_h$ the set of nodes (vertices) on $\tau$, which is also referred to as the skeleton of the (1D-)triangulation $\mathcal{I}_h$.

\subsection{Approximation of surfactants' concentration}

The transport equation of the surfactants' concentration $\Gamma$, Eq. \eqref{eq:surfactant_transport}, can be associated with a high P\'eclet number, which makes it difficult to solve numerically \citep{donea2003finite}.
To this end we chose a locally bound preserving discretisation which can be interpreted as a face-centered finite volume method, see \cite{EYMARD2000713, VIEIRA2020112655}. The surfactants' concentration is approximated at the midpoint (center) of elements in $\mathcal{I}_h$ and additionally at vertices $V \in \mathcal{F}_h$. We define the spaces
\begin{align*}
    G_h = \{ \gamma_h \in \mathbb{P}^0(I), \text{ for all } I \in \mathcal{I}_h \}, \quad \text{and} \quad
    \hat G_h =  \{ \hat \gamma_h \in \mathbb{P}^0(V), \text{ for all } V \in \mathcal{F}_h \},
\end{align*}
where $\mathbb{P}^l(\omega)$, $l\ge 0$ corresponds to the space of polynomials of degree $l$ defined on a given domain $\omega$. 

Considering $\gamma_h$ and $\hat{\gamma}_h$ as the discrete test functions of the spaces $G_h$ and $\hat{G}_h$, we define the following weak formulation for the surfactants' concentration transport equation \eqref{eq:surfactant_transport}:
\
\begin{align}
  \sum_{I \in \mathcal{I}_h} \Big( -\int_{I} \Gamma_h \mathbf{u}_h \bcdot \bnabla \gamma_h dS +
  \int_{{\partial I}} \mathbf{u}_h \bcdot \mathbf{n} \hat{\Gamma_h}\gamma_h ds + \int_{\partial I_\text{out}} \mathbf{u}_h \bcdot \mathbf{n} (\Gamma_h - \hat{\Gamma_h}) {\hat \gamma}_h ds \nonumber  \\
  + D^\text{surf} \dfrac{2}{h^2} \int_{\partial I} (\Gamma_h - \hat{\Gamma}_h)(\gamma_h - \hat{\gamma}_h) ds \Big)= 0 \, .
  \label{eq:surfactant_transport_weak}
\end{align}
\
\noindent Here, the first three terms correspond to the convection, while the latter describes the diffusion. Note, that due to the element-wise constant approximation, the first integral vanishes. The vector $\mathbf{n}$ denotes the outward pointing normal vector of an element $I$ (defined at the vertices of $I$, i.e. tangential to the interface), and $\partial I$ reads as the two end points (vertices) of $I$. An \textit{outflow boundary} $\partial I_{\text{out}}$ is understood as an endpoint where $\mathbf{u}_h \bcdot \mathbf{n} \ge 0$. 
Further note that an integral of the form $\int_{\partial I} \cdot ds$ actually reads as point evaluations at the two end points. 
Testing the equation solely with test functions ${\hat\gamma}_h$, it is easy to see that (for $D^\text{surf} = 0$) the trace approximation ${\hat\Gamma}_h$ corresponds to the upwind-value, thus we consider an upwind stabilised convection formulation. 
As known from finite-volume methods (and lowest order discontinuous Galerkin methods, see \citet{EYMARD2000713, doi:10.1142/13466}), this results in a locally conservative method, and the approximated values of $\Gamma_h$ will stay positive. 
We further want to emphasise the factor $2$ in the diffusive bilinear form which results from considering \textit{finite differences} between the element midpoint (approximated via $\Gamma_h$) and the trace at the boundary ${\hat \Gamma}_h$ (which are apart by $h/2$, where $h$ is the length of the element $I$).

\subsection{Approximation of the flow velocity, pressure, temperature and vapour concentration}

The approximation of the Stokes problem is based on the MINI finite element method (see \cite{boffi2013mixed}), i.e. we approximate the velocity and the pressure in the spaces
\begin{align*}
    V_h = \{\mathbf{u}_h \in \mathbb{P}^3(K)\cap C^0(\Omega): \mathbf{u}_h|_{\partial K} \in \mathbb{P}^1(\partial K) \text{ for all } K \in \mathcal{T}_h\}, \\
    Q_h = \{  p_h \in \mathbb{P}^1 \cap C^0(\Omega)\},
\end{align*}
respectively. Thus, the velocity is element-wise approximated by linear functions including the local element bubble (a cubic polynomial which vanishes at element boundaries $\partial K$). We have chosen this discretisation because numerical experiments showed that it is essential that the (surface) gradient of the velocity (for the MINI-element resulting in an element-wise constant since the bubble part vanishes) matches the approximation space $G_h$. Although other combinations are numerically stable, they resulted in oscillations in the tangential velocity. 

As already denoted in section \ref{sec:procedure}, both the temperature and vapour concentration is discretised by a standard first-order Lagrangian finite element method. 

\section{Dependence on Lewis number}\label{ap:lewis_number}

\begin{figure}
  \centerline{\includegraphics[width=0.7\textwidth]{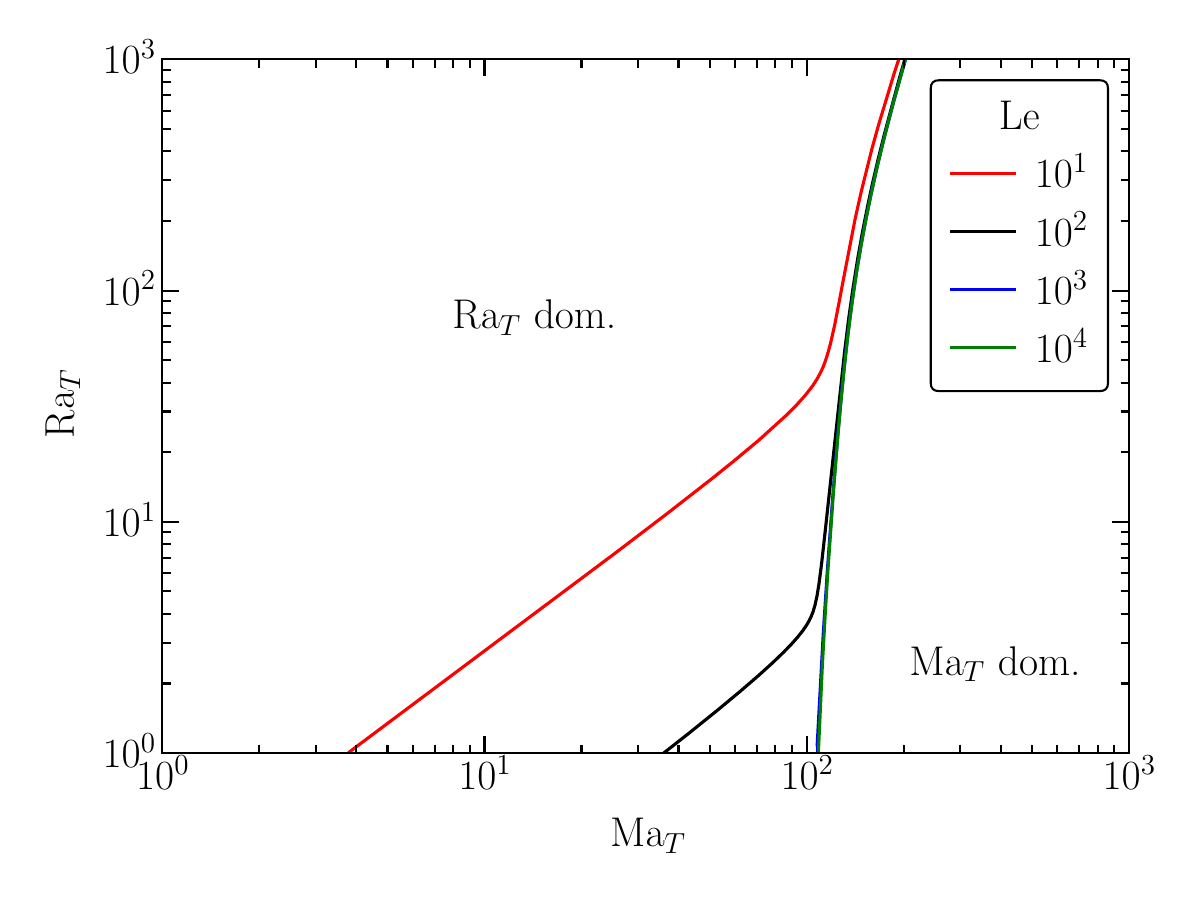}}
  \caption{$\MAT$-$\RAT$ phase diagram for flow direction for a droplet of 150$^\circ$ contact angle and $\MAG$=500, considering different $\operatorname{Le}$ numbers,
  $\operatorname{Le}$=\num{e1} (red), $\operatorname{Le}$=\num{e2} (black, value taken in results of figure \ref{fig:phase_diagram_largerMAG}), $\operatorname{Le}$=\num{e3} (blue) and $\operatorname{Le}$=\num{e4} (green).}
  \label{fig:phase_diagram_Le_number}
\end{figure}

In the results of figure \ref{fig:phase_diagram_largerMAG}, we have considered a fixed surface diffusivity of \SI{e-9}{\meter^2\per\second}, i.e.\ $\operatorname{Le}$ number of $\sim$\num{e2}\, .
Although this assumption is taken, the unknown source of the surfactants makes it difficult to determine a valid $\operatorname{Le}$ number.
Notably, its influence on the flow direction is not expected to be entirely negligible. 
Figure \ref{fig:phase_diagram_Le_number} shows the $\MAT$-$\RAT$ phase diagram for flow direction for a droplet of 150$^\circ$ contact angle and $\MAG$=500, considering different $\operatorname{Le}$ numbers, $\operatorname{Le}$=\num{e1} (red), $\operatorname{Le}$=\num{e2} (black, value taken in results of figure \ref{fig:phase_diagram_largerMAG}), $\operatorname{Le}$=\num{e3} (blue) and $\operatorname{Le}$=\num{e4} (green).

On the one hand, higher diffusivity in the transport equation \eqref{eq:surfactant_transport}, i.e. lower $\operatorname{Le}$, implies a weaker transport of surfactants by interfacial velocity towards the apex.
At the apex, for high contact angles, thermal gradients are more intense. 
If the accumulation of the surfactants at the apex is weaker, the surface tension gradient due to thermal effects will be stronger, therefore leading to a more pronounced thermal Marangoni flow.
On the other hand, a lower diffusivity weakens the thermal Marangoni flow.
However, a large compression of the surfactants leads to reaching a zero-concentration at the contact line for lower $\MAT$.
Thereby, the flow reversal curve happens when $\MAT$ is strong enough to push the surfactants away from the contact line, allowing it to grow and develop a strong flow close to the contact line.

\section{Contact angle influence on axisymmetric flow direction} \label{sec:contact_angle_flow_direction_phase_diagram}

The spatial scaling for the dimensionless equations is based on the volume of the droplet. 
The contact angle is then another parameter that can influence the flow direction. 
We investigate here the impact of \CA\ on the flow direction in the $\RAT$-$\MAT$ phase diagram for droplets of the same volume.
For \CA$<90^\circ$, even though the evaporation rate is higher at the contact line, the temperature at the apex is still lower if we assume the substrate to be perfectly conductive during the evaporation process. 
In such cases, the thermal Marangoni flow will have the same direction as for higher \CA, i.e.\ from the contact line towards the apex.

\begin{figure}
    \centerline{\includegraphics[width=1\textwidth]{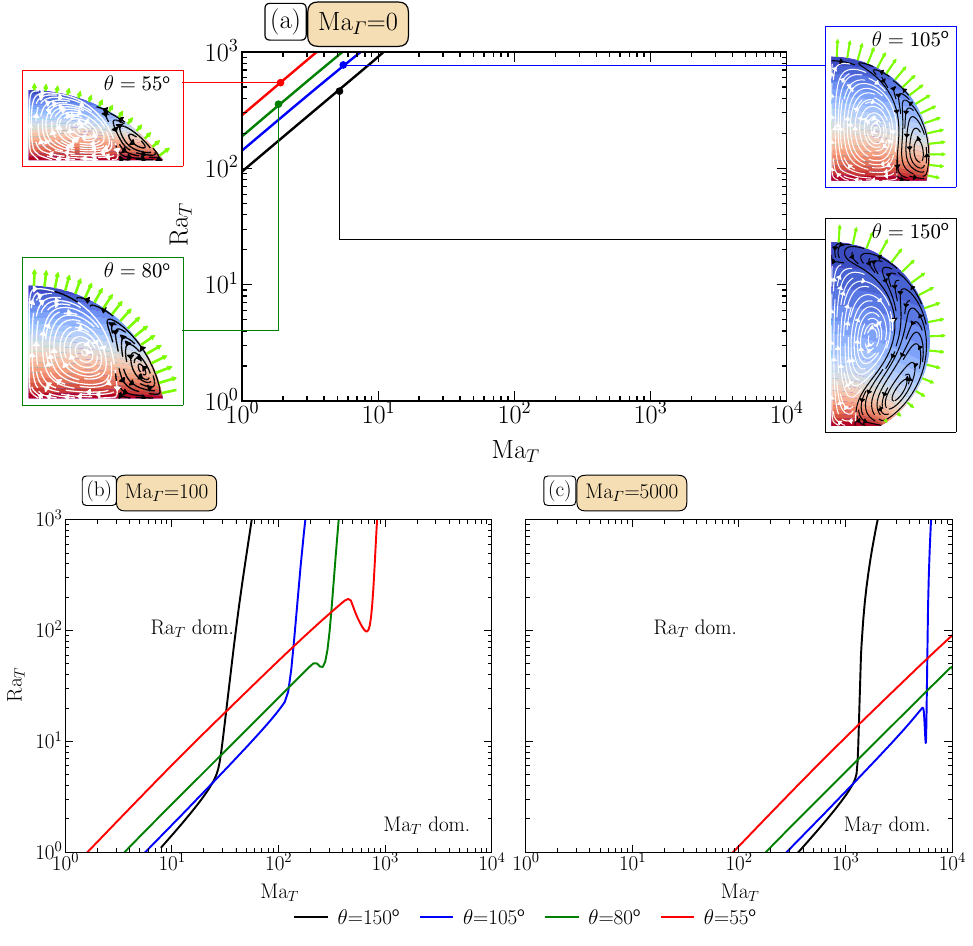}}
    \caption{Flow reversal curves (i.e.\ $\psi=50\%$) in the $\MAT$-$\RAT$ phase diagram for \CA{} = 150$^\circ$ (black), 105$^\circ$ (blue), 80$^\circ$ (green) and 55$^\circ$ (red), and $\MAG$=0 (a), 100 (b) and 5000 (c). 
    At the top, the insets show the flow scenario in the droplet for parameters directly on the reversal curve, for each respective contact angle.}
    \label{fig:flow_reversal_largerMAG_90}
\end{figure}

Figure \ref{fig:flow_reversal_largerMAG_90} illustrates the $\psi^+=50\%$ curves. 
The curves are shown for contact angles of $55^\circ$ (red), $80^\circ$ (green), $105^\circ$ (blue), and the reference $150^\circ$ (black), for $\MAG$ of 0 (a), 100 (b), and 5000 (c).
When $\MAG = 0$, for the flow to be from the contact line towards the apex, lower \CA{} requires $\RAT$ to be larger, for a fixed $\MAT$. 
This is expected since the temperature gradient is directly related to the height and contact radius of the droplet, which are respectively smaller and larger for droplets of the same volume with lower \CA.
Lower thermal Rayleigh effects are therefore expected for the latter case.

For $\MAG \ne 0$ and low $\MAT$, the curves are nearly linear, and for a fixed $\RAT$, lower \CA\ require lower $\MAT$ for the flow to be dominated by thermocapillary forces. 
As $\MAT$ increases, the $\psi^+=50\%$ curves change slope, as discussed in \S\ref{sec:zero_surfactants}. 
This change is attributed to the surfactant concentration at the contact line reaching zero, which occurs at lower $\MAT$ for higher \CA{}. 
Higher \CA{} values imply larger temperature gradients within the droplet, leading to stronger thermal Marangoni flows. 
Consequently, the tangential velocity at the interface is higher for higher \CA, causing surfactants to be advected away from the contact line more easily and altering the $\psi^+=50\%$ slope at lower $\MAT$ values.

\bibliographystyle{myjfm}
\bibliography{resubmission}

\end{document}